\documentclass[trackchanges,twocolumn]{aastex631}

\usepackage{graphicx} 
\usepackage{multirow}
\usepackage{footmisc}
\usepackage{makecell}
\usepackage{amsmath}
\usepackage{multirow}
\usepackage{ulem}

\newcommand{\Msun}{\hbox{$M_\sun$}}
\newcommand{\AU}{\hbox{$\text{AU}$}}

\newcommand{\RB}{\hbox{$\Delta_{\text{RB}}$}}
\newcommand{\RO}{\hbox{$\Delta_{\text{R}\sun}$}}
\newcommand{\EB}{\hbox{$\Delta_{\text{EB}}$}}
\newcommand{\SB}{\hbox{$\Delta_{\text{SB}}$}}

\newcommand{\KB}{\hbox{$\Delta_{\text{KB}}$}}
\newcommand{\SR}{\hbox{$\Delta_{\text{SR}}$}}
\newcommand{\AOP}{\hbox{$\Delta_{\text{AOP}}$}}
\newcommand{\OP}{\hbox{$\Delta_{\text{OP}}$}}

\newcommand{\taBB}{\hbox{$t_\text{a,BB}$}}
\newcommand{\tBB}{\hbox{$t_\text{BB}$}}

\newcommand{\kms}{{km~s\textsuperscript{-1}}}

\def\BB {{binary barycenter}}
\def\SSB {{solar system barycenter}}

\defcitealias{DD1}{D\&D~I}
\defcitealias{DD2}{D\&D~II}

\received{July 15, 2022}
\revised{March 22, 2023}
\accepted{April 5, 2023}
\published{July 14, 2023}
\submitjournal{ApJ}

\shorttitle{Analytic PN Models of Orbits around Black Holes}
\shortauthors{Hyman et al.}

\begin{document}

\title{Analytic Post-Newtonian Astrometric and Spectroscopic Models of Orbits around Black Holes}

\correspondingauthor{S\'oley Hyman}
\email{soleyhyman@arizona.edu}

\author[0000-0002-6036-1858]{S\'oley \'O. Hyman}
\affiliation{Steward Observatory and Department of Astronomy,\\University of Arizona, 933 N. Cherry Ave., Tucson, AZ 85721, USA}
\author[0000-0003-1035-3240]{Dimitrios Psaltis}
\affiliation{Steward Observatory and Department of Astronomy,\\University of Arizona, 933 N. Cherry Ave., Tucson, AZ 85721, USA}
\affiliation{School of Physics, Georgia Institute of Technology, North Avenue, Atlanta, GA, 30332}
\author[0000-0003-4413-1523]{Feryal \"Ozel}
\affiliation{Steward Observatory and Department of Astronomy,\\University of Arizona, 933 N. Cherry Ave., Tucson, AZ 85721, USA}
\affiliation{School of Physics, Georgia Institute of Technology, North Avenue, Atlanta, GA, 30332}
%

\begin{abstract}
Observations of the S stars, the cluster of young stars in the inner 0.1~pc of the Galactic center, have been crucial in providing conclusive evidence for a supermassive black hole at the center of our galaxy. Since some of the stars have orbits less than that of a typical human lifetime, it is possible to observe multiple orbits and test the weak-field regime of general relativity. Current calculations of S-star orbits require relatively slow and expensive computations in order to perform numerical integrations for the position and momentum of each star at each observing time. In this paper, we present a computationally efficient, first-order post-Newtonian model for the astrometric and spectroscopic data gathered for the S stars. We find that future, 30-m class telescopes -- and potentially even current large telescopes with very high spectroscopic resolution -- may be able to detect the Shapiro effect for an S star in the next decade or so.
\end{abstract}

\keywords{general relativity, two-body problem, orbital motion, supermassive black holes, Galactic center}

\section{Introduction}\label{sec:intro}
Supermassive black holes are located at the center of most large galaxies \citep[e.g.,][]{1995ARA&A..33..581K,2013ARA&A..51..511K} and provide a unique environment for probing the effects of general relativity (GR). The Milky Way contains its own central black hole, Sgr~A* \citep{1954Natur.173..985M,1971Natur.233..112D,1989IAUS..136..527L,1993Natur.362...38L,1994ASIC..445..403B,1994RPPh...57..417G,1998ApJ...509..678G,1999JApA...20..187E}, which is surrounded by a cluster of young stars \citep[referred to as S stars;][]{1996Natur.383..415E,1997MNRAS.284..576E,1998ApJ...509..678G,1998IAUS..184..421G,1999A&A...352L..22E}. Measuring their orbits have helped measure the ratio of the mass-to-distance ratio of Sgr~A* \citep{1996Natur.383..415E,1996ApJ...472..153G,1997MNRAS.291..219G,1997MNRAS.284..576E,2000Natur.407..349G,2008ApJ...689.1044G,2013ApJ...779L...6D,2016ApJ...830...17B}.

One of the closest stars to Sgr~A* is S0-2 \citep[also known as S2;][]{2002Natur.419..694S,2003ApJ...586L.127G,2009ApJ...707L.114G}, which has an orbital period of around 16~years and eccentricity of 0.88. Long-term studies of its orbit led to the first detection of gravitational effects during its 2018 periapsis -- namely gravitational redshift \citep{2018A&A...615L..15G,2019A&A...625L..10G,2019Sci...365..664D,2021A&A...647A..59G,2022A&A...657L..12G} and Schwarzschild precession \citep{2020A&A...636L...5G} -- in an S-star orbit. As both photometric and spectroscopic sensitivities improve and shorter-period S-star candidates are identified \citep[e.g.,][]{2020ApJ...889...61P,2020ApJ...899...50P}, additional tools are needed to analyze and detect higher-order general relativistic effects, such as the Shapiro delay, and additional precession due to the frame dragging and quadrupole moment of the spacetime \citep[e.g.,][]{1999ApJ...514..388W,2004AAS...204.1702W,2008ApJ...674L..25W,2010PhRvD..81f2002M,2010ApJ...720.1303A,2014MNRAS.444.3780A, 2016ApJ...818..121P,2017A&A...608A..60G,2018MNRAS.476.3600W}.

Current modeling of S-star orbits involves integrating numerically the general relativistic (GR) equations of motion for each time step \citep[e.g.,][]{2017ApJ...837...30G,2019Sci...365..664D}. This method requires integrating across the span of observations using very small time steps to avoid error buildup and results in slow, expensive computations. Furthermore, the computational cost of such numerical calculations increases rapidly when using the orbits of multiple stars to jointly constrain the shared properties of the system (e.g., the black hole mass), since this involves simultaneously solving the geodesic equations for each time step for each star. This approach could become prohibitive when searching the multi-dimensional orbital parameter space with a statistical sampling algorithm, such as a Markov Chain Monte Carlo (MCMC) to obtain optimal solutions and quantify uncertainties in orbital parameters.

A simplification to these calculations can be introduced because of the fact that S-star orbits have pericenter distances that range from 1,400~Sgr~A*~Schwarzschild~radii to values that are larger by orders of magnitude \citep{2017ApJ...837...30G}. At such distances, the orbits can be modeled as primarily Keplerian orbits with small corrections caused by GR effects. A framework for describing this behavior is through a semi-analytic post-Newtonian (PN) model, which uses traditional Keplerian orbital equations with additional terms up to some order in $v/c$, derived from GR equations. 

Damour \& Deruelle \citeyearpar[referred to hereafter as D\&D~I and D\&D~II]{DD1,DD2} obtained an elegant analytic solution to the two-body problem in the first post-Newtonian order (1PN), which incorporates a variety of relativistic effects. The timing model developed by \citetalias{DD2} has been the workhorse for the pulsar community for many years in detecting relativistic effects \citep[e.g.,][]{Tempo2006}. It was further expanded to include second-order post-Newtonian (2PN) terms \citep{1988NCimB.101..127D,1995CQGra..12..983W}.

While the \citetalias{DD2} analytic solution has been implemented in a timing model for fitting pulsar arrival times, the same model is not readily applicable for fitting astrometric and spectroscopic data of stars. This is because the latter relies on the Doppler shifts of emission lines in the stellar spectra as opposed to time intervals between pulses. The beauty of the analytic \citetalias{DD2} timing model, however, makes it possible to derive the line-of-sight velocities that correspond to a variety of time delays.

In this paper, we use the framework of the \citetalias{DD1} and \citetalias{DD2} 1PN two-body model (in harmonic coordinates) to derive a new analytic astrometric and spectroscopic model that incorporates the relativistic and astrophysical effects relevant to modeling S-star orbits. 

In addition to computational efficiency, there are significant scientific advantages to our approach. In the numerical approach, all relativistic effects of the same post-Newtonian order that are embedded in the geodesic equations are reported as a single observable (i.e., the position in the sky or spectroscopic line shift). In principle, the magnitudes of the individual effects can be disentangled by exploring the differences in the numerical solutions with and without each effect~\citep{2017A&A...608A..60G}. In contrast, a post-Newtonian analytic approach, such as our model, allows for calculating directly the characteristic “fingerprint” of each effect on the observables separately from the others (see Section~\ref{sec:spec-effecs}), while providing a direct analytic handle of the dependence of each effect on the various parameters of the system.

The following sections present the two-body orbital equations (\S\ref{sec:orbital-model}), the projection of those equations to the plane of the sky (\S\ref{sec:astromet}), and the equations for spectroscopic effects (\S\ref{sec:spec-model}). We discuss the implications of the model for S-star observations in Section~\ref{sec:spec-effecs}. Unless otherwise indicated, we use geometrized units, i.e., $G=c=1$, where $G$ is the gravitational constant and $c$ is the speed of light.

\section{Orbital Model}\label{sec:orbital-model}

As discussed in Section~\ref{sec:intro}, our model has three main components: the two-body orbital model, the astrometric model, and the spectroscopic model. Since we want to use our model to be able to fit observations of the S stars, we first identify what parameters are the observables. With telescopes, we are able to observe the projected positions of the stars, i.e., right ascension (R.A., $\alpha$) and declination (Decl., $\delta$), and the radial velocities derived from their spectra, 
\begin{equation}\label{eq:los-velocity}
    v_z(t_\text{obs}) = \frac{\Delta\lambda(t_\text{obs})}{\lambda_0}\;,
\end{equation}
where $\lambda_0$ is the rest-frame wavelength of the stellar absorption or emission line used for measuring radial velocities and $\Delta\lambda$ is the shift at time $t_\text{obs}$ between observed and rest-frame wavelengths.

Four free parameters -- orbital period ($P$), total mass of the system ($M$), mass ratio of the two bodies ($q$), and radial eccentricity ($e_R$) govern the shape, period, and rate of precession of the two-body orbits. 

The orientations of the orbits with respect to Earth determine the two-dimensional motions in the sky (i.e., the astrometric model) and the line-of-sight motions, which we can derive through spectroscopy. The three orientation angles -- inclination ($i$), argument of ascending node ($\Omega$), initial argument of periapsis ($\omega_0$), and initial time of periapsis (or epoch of position, $t_0$) -- are free parameters for our model. We must also fit the projected and line-of-sight proper motions of the Galactic center with respect to Earth ($\mu_\alpha$, $\mu_\delta$, $\mu_\parallel$).

\begin{figure}[htb!]
\centering
\includegraphics[width=0.45\textwidth]{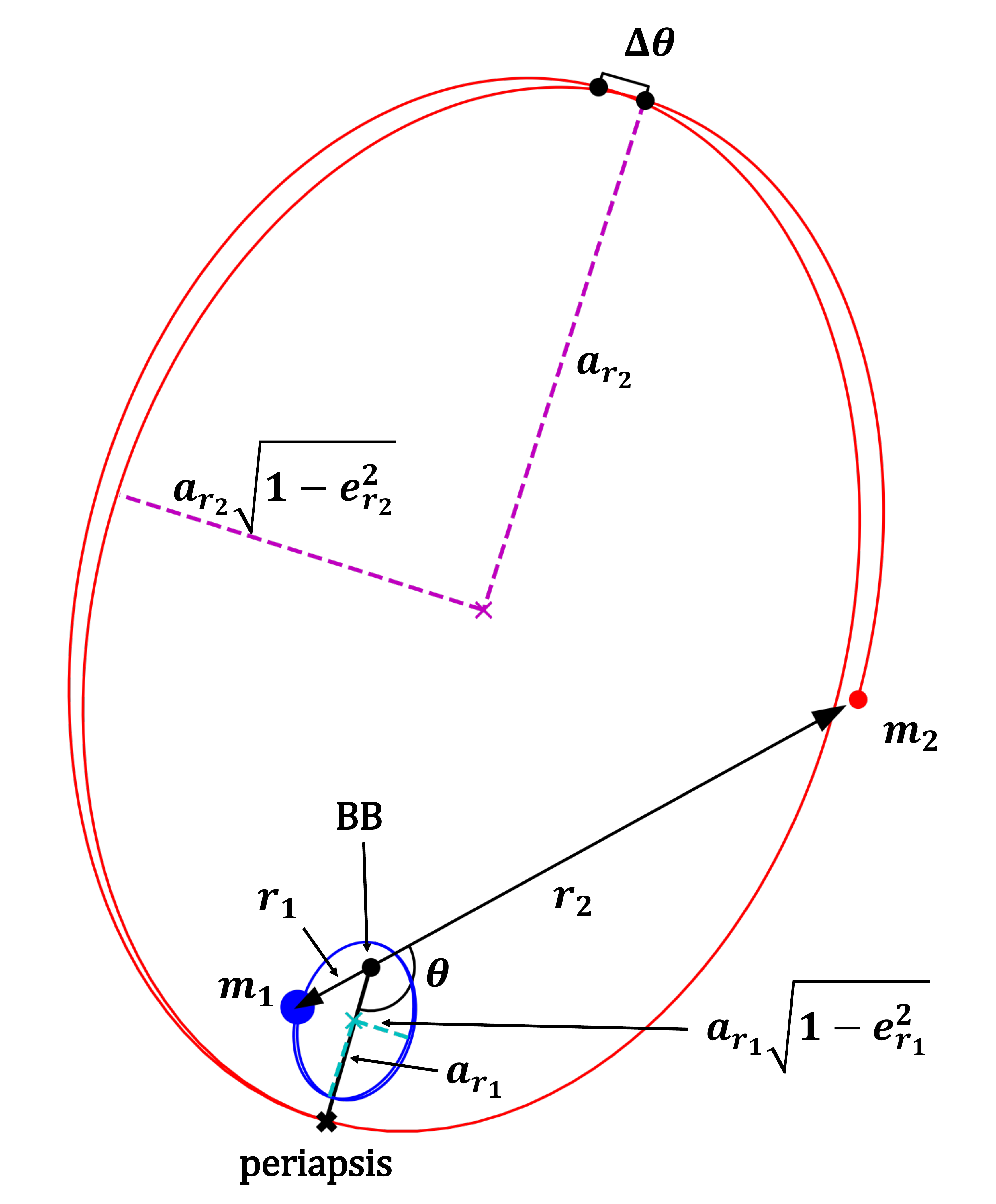}
\caption{The binary orbital parameters in the barycenter frame of the system. The large red, precessing ellipse shows the motion of the smaller mass $m_2$ over roughly two periods. Similarly, the smaller blue ellipse shows the motion of the larger mass $m_1$ over the same amount of time. Dotted magenta lines mark the semimajor and semiminor axes of the ellipses for both objects around the \BB, which is indicated by a black dot. Arrows between the \BB\ and the two masses denote the distances $r_1$ and $r_2$ between the \BB\ and orbiting objects. A black \texttt{x} marks the closest approach that the smaller mass $m_2$ makes to the larger mass $m_1$, which precesses by an angle $\Delta\theta$ for each orbit. The position angle $\theta$ is the angle formed between the periapsis point and the location of one of the masses, which are offset from each other by 180\degr.}
\label{fig:orbital-params}
\end{figure}

Nineteen additional quantities (introduced in the following sections) are derived parameters that we calculate from the observed or free parameters. Table~\ref{table:params_symbols} in Appendix~\ref{appendix:parameters} lists all the parameters introduced in this paper with appropriate references and units. Figure~\ref{fig:orbital-params} illustrates how the binary system orbital parameters relate to each other.

In this section, we present the orbital equations and parameters in geometrized units.

\subsection{Coordinate Systems}\label{subsec:coords}
\begin{figure}[htb!]
\centering
\includegraphics[width=0.45\textwidth]{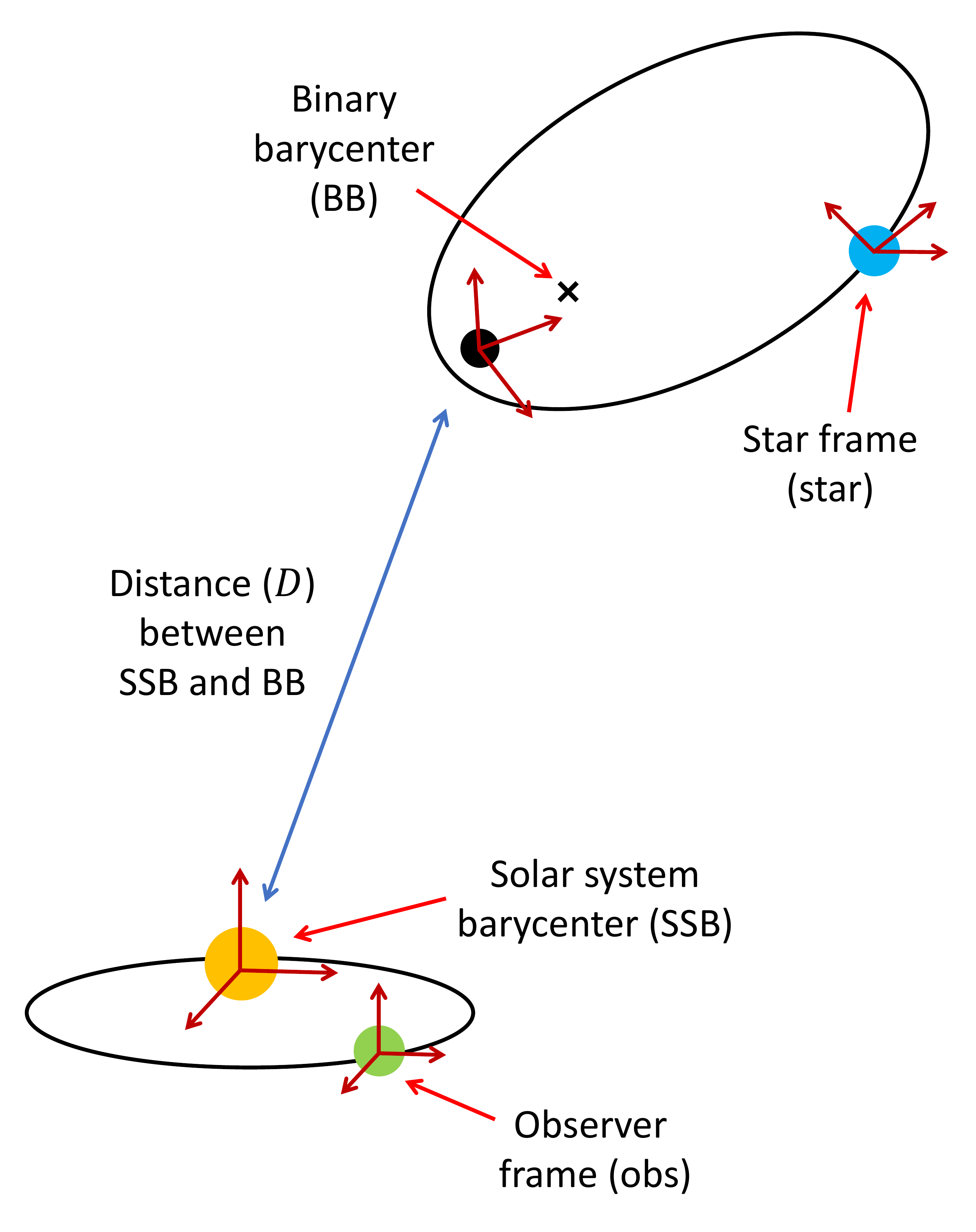}
\caption{The relationship between the observer frame, the \SSB, the \BB, and the star frame, as defined in this paper. Distances and object sizes have been rescaled to show effect. The red axes for each reference frame are arbitrary to show how coordinate systems may vary from frame to frame depending on orientation.}
\label{fig:ref-frames}
\end{figure}

We use four main coordinate systems/reference frames (Figure~\ref{fig:ref-frames}), all of which are described in \citet{Tempo2006}. These are: the star reference frame (denoted by the subscript ``star"), the binary barycenter (BB), the \SSB \ (SSB), and the observer reference frame (``obs"). The different reference frames are necessary to define the various time delays discussed in Section~\ref{sec:spec-model}. All celestial sky coordinates are given in the International Celestial Reference System \citep[ICRS,][]{2015HiA....16..227L}.

The times (or ``clocks") we use in this paper are directly related to the four reference frames. We define the time of emission as measured at the stellar location as $t_\text{star}$, the same time as measured by an observer at the \BB\ as \tBB, the time of arrival at the \BB\ as \taBB, the time of arrival at the \SSB\ as $t_\text{SSB}$, and the time of arrival recorded by the observer on Earth as $t_\text{obs}$.

The relation between the observer light-arrival time $t_\text{obs}$ and the star light-emission time $t_\text{star}$ (used for the time-dependent orbital equations in Section~\ref{subsec:time-dep-eqs}) is the star-frame emission time plus the sum of all time delays due to binary system motion, solar system motion, and motion between the \BB\ and \SSB/observer, i.e.,
\begin{equation}\label{eq:tobs-tstar}
    \begin{array}{llr}
    t_\text{obs} = & t_\text{star} &\\
    &  +\RB + \EB + \SB & \text{[binary effects]}\\
    &+ \KB & \text{[parallax effects]}\\
    &+ \RO\;. & \text{[solar system effects]}
    \end{array}
\end{equation}
In our model, the total binary system time delay includes the binary Roemer delay (\RB), binary Einstein delay (\EB), and binary Shapiro delay (\SB). The total solar-system-related time delay is the Earth Roemer delay (\RO). The interstellar time delay comprises time delays due to parallax and proper motion, which is simply the Kopeikin effect (\KB). We provide explicit formulae for the time delays in Section~\ref{sec:spec-model}.

Additional relations between the other time variables are as follows. The binary Einstein delay relates the star emission time to the time as measured at the \BB\ by an inertial observer as $\tBB = t_\text{star} + \EB$. The time of arrival at the \BB\ relates to the star emission time via the binary effects, i.e., $\taBB = t_\text{star} + \RB + \EB + \SB$. Similarly, the observer time of arrival relates to the \SSB\ time of arrival via the solar system effects, i.e., $t_\text{obs} = t_\text{SSB} + \RO$. Since the spectroscopic timing model is less sensitive than pulsar timing, we neglect interstellar delays (such as dispersion). As a result, the \SSB\ time of arrival and \BB\ time of arrival are related by a constant offset, which we set to zero without loss of generality.

\subsection{Mass Measures}\label{subsec:mass-eqs}
Three of the key values in describing a two-body orbital system are the total system mass and the individual masses of the two objects. We denote the component masses by $m_1$ and $m_2$, such that $m_2 \leq m_1$. In the case of modeling stars orbiting a supermassive black hole, $m_1$ is the mass of the black hole and $m_2$ is the mass of the star. The total mass of the system is $M=m_1 + m_2$.
\begin{subequations}
In classical, two-body orbits, the mass ratio of the two objects,
    \begin{equation}\label{eq:q}
       q = \frac{m_2}{m_1} \;,
    \end{equation}
is an important parameter that we can use to rewrite the reduced mass,
    \begin{equation}\label{eq:mu}
       \mu = \frac{m_1 m_2}{m_1 + m_2} = \frac{m_2}{1+q}\;,
    \end{equation}
and dimensionless reduced mass ($\nu$) as
    \begin{equation}\label{eq:nu}
       \nu = \frac{\mu}{m_1 + m_2} = \frac{q}{(1+q)^2} \;.
    \end{equation}
Note that in Sgr~A*-S-star systems, the mass ratios can be $q\sim~10^{-6}-10^{-5}$. In these cases, one can take the limit of $q\ll1$, but in this paper we leave the full expressions for generality.
\end{subequations}

\subsection{Period, Semimajor Axis, and Mean Motion}\label{subsec:psm-eqs}
The easiest, most direct property to measure is the orbital period, $P$. With the total system mass and orbital period, we use the 1PN version of Kepler's law \citep[][their Eq.~8.6]{1998PhRvD..58l4002B} to calculate the center-of-mass semimajor axis, implicitly, via
\begin{subequations}
    \begin{equation}\label{eq:aR}
        P^2 = \frac{M}{a_R^3}\left[1+(-3+\nu)\frac{M}{a_R}\right]
    \end{equation}

Another useful related quantity is the mean motion, the constant angular speed needed for an object to complete an equivalent circular orbit. It relates to the inverse period as
     \begin{equation}\label{eq:n}
        n = \frac{2\pi}{P}\;.
    \end{equation}
\end{subequations}

\subsection{Energy and Momentum}\label{subsec:momen-eqs}
In addition to calculating the semimajor axis $a_R$ from $P$, we also fit for the eccentricity $e_R$. In combination with the system mass $M$ and the dimensionless reduced mass $\nu$, these two parameters set the orbital behavior and we use them to calculate the energy and angular momentum of the system. Here, the total energy is
\begin{subequations}
    \begin{equation}\label{eq:E}
        E = \frac{2 M}{M (\nu-7) - 4a_R}\;
    \end{equation}
and total angular momentum is
    \begin{equation}\label{eq:J}
        J = M \Bigg\{ \frac{-1 + e_R^2 - E[2 + 5E (\nu-3)](\nu-6)}{E[2 + 5E (\nu-3)]} \Bigg\}^{1/2}\;.
    \end{equation}

We also define the quantity $K$, which is the general relativistic correction to the total angular momentum of the system, rewritten in natural units from \citetalias{DD1} Equation~(4.14) and given by
    \begin{equation}\label{eq:K}
       K = \frac{J}{\sqrt{J^2 - 6M^2}} \;.
    \end{equation}
This is a particularly important quantity, as the value of $K$ is what governs the orbital precession ,
    \begin{equation}\label{eq:precess}
        \Delta\theta = 2\pi(K-1)\;,
    \end{equation}
where $\Delta\theta$ is the angle the orbit precesses in each period \citepalias{DD1}.
\end{subequations}

\subsection{Other Eccentricities}\label{subsec:sem-eqs}
While Keplerian orbits have only one effective eccentricity, $e_R$, in GR, there are additional eccentricities that result from the curved spacetime \citepalias[their Eqs.~3.6b-c,~4.13]{DD1}. In the 1PN model, these are the time eccentricity \citepalias[$e_t$;][their Eq.~3.6c]{DD1},
\begin{subequations}
    \begin{equation}\label{eq:et}
        e_t = e_R \left[1+\frac{M}{a_R} \left(4-\frac{3}{2}\nu\right)\right],
    \end{equation}
and the angular eccentricity \citepalias[$e_\theta$;][their Eq.~4.13]{DD1}, which in natural units is:
    \begin{equation}\label{eq:etheta}
        e_{\theta} = e_R \left(1+\frac{\mu}{2a_r}\right)\;.
    \end{equation}
In some cases, the differences between the eccentricities are negligible and all eccentricity expressions give comparable answers (see Section~\ref{sec:spec-model} for examples). The radial eccentricity can be determined readily from observational astrometric data.
\end{subequations}

\subsection{Individual Objects Parameters}\label{subsec:obj-eqs}
As noted in Section~\ref{subsec:psm-eqs}, the semimajor axis and radial eccentricity defined here are with respect to the center of mass of the system. Since observations of the S stars result in tracking the orbits of the individual stars, we need the derived parameters (i.e., semimajor axis and radial eccentricity) that give the orbital shapes of both objects in a two-body system, which we can derive from the corresponding effective one-body parameters (i.e., $a_R$ and $e_R$) and the mass ratio ($q$).

For the more massive of the two bodies, $m_1$, the semimajor axis of its respective orbit around the \BB\ is
\begin{subequations}
    \begin{equation}\label{eq:ar1}
        a_{r_1} = a_R \frac{q}{1+q}\;,
    \end{equation}
and the radial eccentricity is
    \begin{equation}\label{eq:er1}
        e_{r_1} = e_R \left[1 + \frac{m_1 (q-1)}{2a_R (q+1)} \right]
    \end{equation}
\citepalias[][their Eqs.~6.3a-b, in natural units and mass ratio $q$]{DD1}.

Similarly, the less massive of the two bodies, $m_2$, follows an orbit around the \BB\ with a semimajor axis of
    \begin{equation}\label{eq:ar2}
        a_{r_2} = a_R \frac{1}{1+q}
    \end{equation}
and a radial eccentricity of    
    \begin{equation}\label{eq:er2}
        e_{r_2} = e_R \left[1 - \frac{m_2 (q-1)}{2a_R (q+1)} \right]
    \end{equation}
\citepalias[][their Eqs.~6.3a-b, in natural units and mass ratio $q$]{DD1}.
\end{subequations}

\subsection{Time-dependent Orbital Motion}\label{subsec:time-dep-eqs}
In Sections~\ref{subsec:mass-eqs} through \ref{subsec:obj-eqs}, we presented the equations necessary for calculating many of the derived parameters in the model. In this section, we use those parameters to obtain the time-dependent orbital motion of the individual objects and the binary barycenter. 

The heart of this time dependence comes from Kepler's equation, which relates time eccentricity ($e_t$), mean motion ($n$), mean anomaly ($u$), star time of emission ($t_\text{star}$), and epoch of position ($t_0$) via \citetalias{DD1} (their Eq.~3.3),
\begin{subequations}
    \begin{equation}\label{eq:kepler}
        u-e_t \sin(u) - n (t_\text{star} - t_0) = 0\;.
    \end{equation}

The mean anomaly, which is the angle between the periapsis of an orbit and another position in the orbit at some time, is crucial for calculating the other values in the polar orbital equations (i.e., radius and angle). The previous equation does not have an analytical solution for $u$ but can be solved using a fast algorithm, such as the Newton-Raphson method.
Note that, unlike the time-dependent equations in the astrometric model (\S~\ref{sec:astromet}), which use the observer time $t_\text{obs}$, Kepler's equation is evaluated at the star time of emission $t_\text{star}$. This difference, described by Equation~\ref{eq:tobs-tstar}, takes into account the vacuum retardation effect for the astrometric model.

The time-dependent distances of the two orbiting bodies from the barycenter, as given in \citetalias{DD1} (their Eqs.~7.1d-e), are
    \begin{equation}\label{eq:r1}
        r_1 = a_{r_1} (1-e_{r_1} \cos u)\;,
    \end{equation}
and
    \begin{equation}\label{eq:r2}
        r_2 = a_{r_2} (1-e_{r_2} \cos u)\;.
    \end{equation}

Calculating the position angle $\theta$ of the orbiting objects is somewhat more complicated. \citetalias{DD1} present the rather straightforward equation (their Eq.~4.11b)
    \begin{equation}
        \theta = \theta_0 + K \times 2\arctan\left[\left(\frac{1+e_{\theta}}{1-e_{\theta}}\right)^{1/2} \tan\left(\frac{u}{2}\right)\right]\;,
    \end{equation}
but using this in computations requires special care because the evaluation of the term $\tan\left(u/2\right)$ results in floating-point errors for the calculated value of $\theta$ around the asymptotes (i.e., $u=x\pi$ for odd $x$). Instead, we use a series expansion \citepalias[][their Eq.~17d]{DD2} of this equation \citepalias[][their Eq.~4.11a]{DD1}, which avoids these computational issues:
    \begin{equation}\label{eq:theta}
        \theta = \theta_0 + K \times A_e(u)\;,
    \end{equation}
where
    \begin{equation}
        A_e(u) = u + 2\sum_{j=1}^{\infty} \frac{1}{j} \left[\frac{e_{\theta}}{1+(1-e_{\theta}^2)^{1/2}}\right]^j \sin(ju)\;.
    \end{equation}

We performed convergence tests of the series $A_e(u)$ at different values of angular eccentricity ($e_\theta$) and determined that only a small number of terms is needed for necessary computational accuracy, e.g., 30 terms for fractional errors of less than $10^{-8}$.

\begin{figure}[htb!]
\centering
\includegraphics[trim=0cm 0.25cm 0cm 0.75cm, clip, width=\columnwidth]{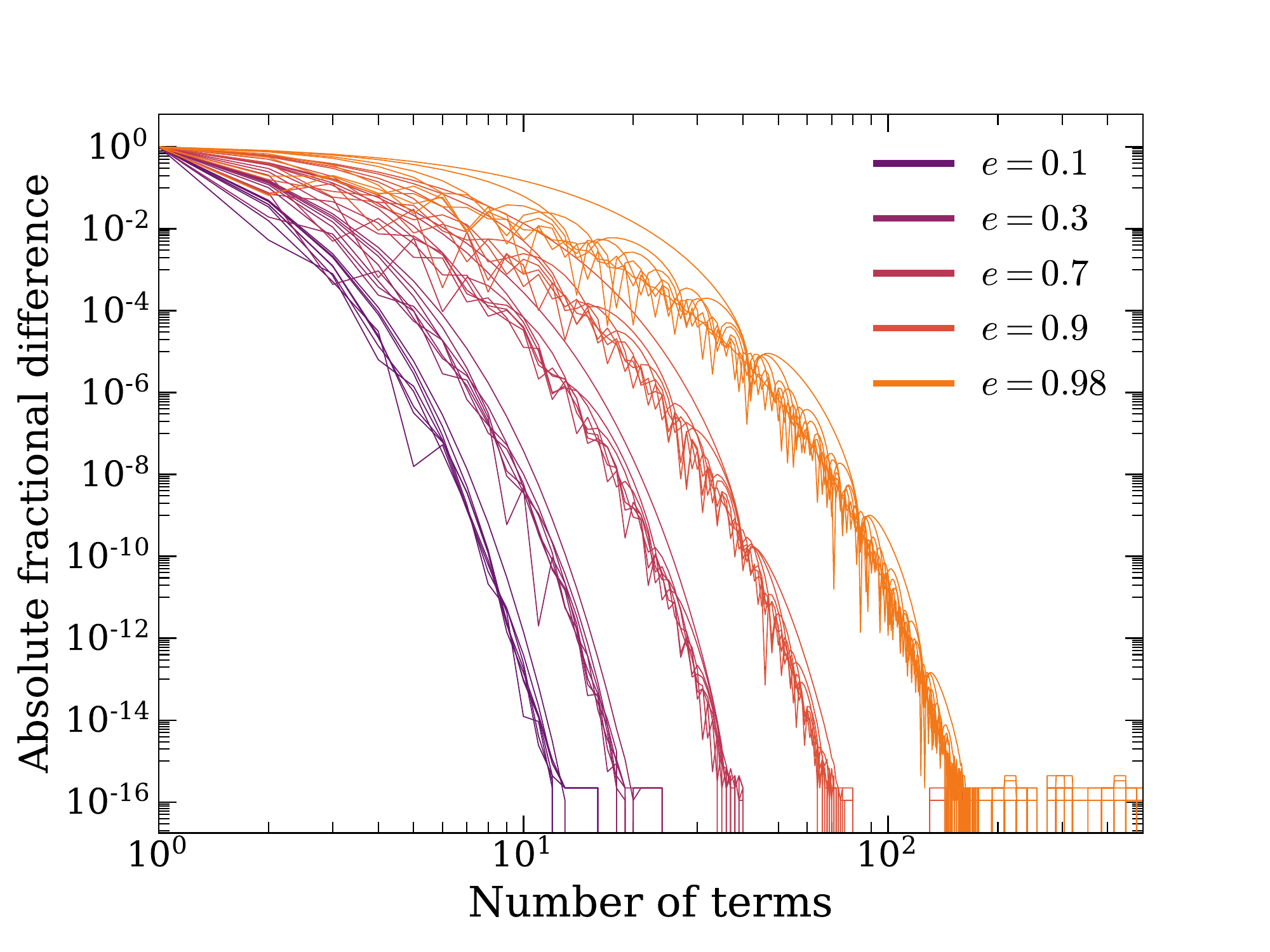}
\caption{Convergence plots for the calculation of the position angle of orbiting objects ($\theta$) with eccentricities of 0.1, 0.3, 0.5, 0.7, 0.9, and 0.98 for a variety of different orbital phases. With the exception of very high eccentricities, 30 terms are typically sufficient for errors of less than $10^{-8}$.}
\label{fig:theta-convergence}
\end{figure}

Much like the position angle $\theta$, the argument of periapsis also precesses due to GR effects (Equation~\ref{eq:K}). Given the initial argument of periapsis, the time-dependent argument of periapsis (in terms of the mean anomaly, $u$) is
\begin{equation}\label{eq:omega_tdep}
    \omega(u) = \omega_0 + (K-1) A_e(u)\;.
\end{equation}
\end{subequations}

\section{Astrometric Model}\label{sec:astromet}
In the previous section (\S\ref{sec:orbital-model}), the 1PN orbital equations are expressed with respect to the binary barycenter. In order to model observations, however, we must transform them to a frame with respect to the plane of the sky, i.e., in terms of right ascension ($\alpha$) and declination ($\delta$). 

We make a transformation from \BB\ motions to projected sky motions by first converting the polar binary barycentric frame ($r_1$, $r_2$, $\theta$) to a Cartesian binary barycentric frame ($x$, $y$) as is typically done, i.e., 
\begin{subequations}
\begin{equation}\label{eq:r1-to-x1}
    x_1=r_1 \cos\theta\;,
\end{equation}
\begin{equation}\label{eq:r1-to-y1}
    y_1=r_1 \sin\theta\;,
\end{equation}
\begin{equation}\label{eq:r2-to-x2}
    x_2=r_2 \cos(\theta+\pi)\;,
\end{equation}
and
\begin{equation}\label{eq:r2-to-y2}
    y_2=r_2 \sin(\theta+\pi)\;.
\end{equation}

We transform the positions from the Cartesian binary barycentric frame to the plane of the sky using using three angles: inclination ($i$), longitude of the ascending node ($\Omega$), and argument of periapsis ($\omega_0$). The inclination describes the tilt of the orbit with respect to the observer, while the longitude of the ascending node is the rotation of the location of the ascending node (i.e., where the orbit intersects with the reference plane) with respect to the center of mass. Similarly, the argument of periapsis is the rotation of the object point of closest approach to the center of mass in the orbital plane. See Figure~\ref{fig:angles} for a graphic depiction. 

We use the Thiele-Innes constants ($A, B, C, F, G, H$) defined by Equations (20)-(25) in \citet{2019AJ....158....4O} to describe this rotation:
\begin{equation}\label{eq:TH-A}
    A = \cos\Omega~\cos\omega - \sin\Omega~\sin\omega~\cos i\;,
\end{equation}
\begin{equation}\label{eq:TH-B}
    B = \sin\Omega~\cos\omega - \cos\Omega~\sin\omega~\cos i\;,
\end{equation}
\begin{equation}\label{eq:TH-C}
    C = \sin\omega~\sin i\;,
\end{equation}
\begin{equation}\label{eq:TH-F}
    F = -\cos\Omega~\sin\omega - \sin\Omega~\cos\omega~\cos i\;,
\end{equation}
\begin{equation}\label{eq:TH-G}
    G = -\sin\Omega~\sin\omega - \cos\Omega~\cos\omega~\cos i\;,
\end{equation}
and
\begin{equation}\label{eq:TH-H}
    H = \cos\omega~\sin i\;.
\end{equation}

\begin{figure}[thb!]
\centering
\includegraphics[width=0.45\textwidth]{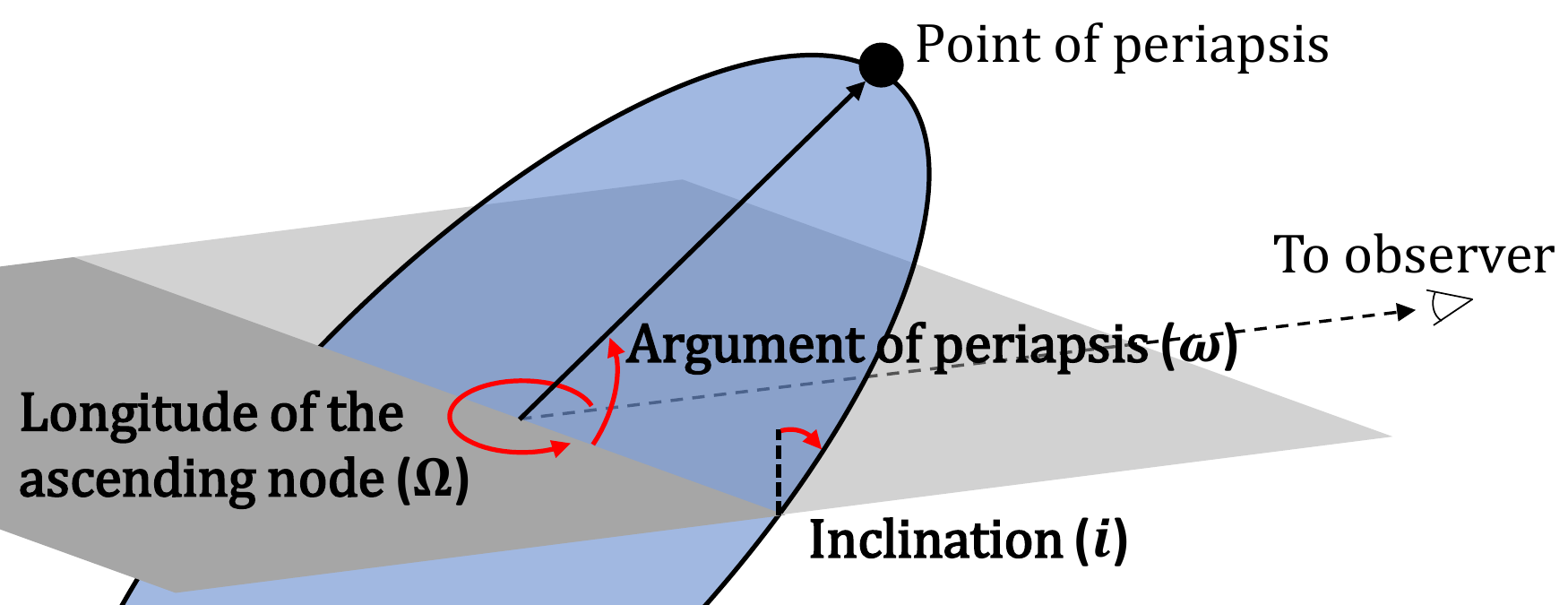}
\caption{Diagram showing the relationship of the three angles used in the astrometric model: inclination ($i$), longitude of the ascending node ($\Omega$), and argument of periapsis ($\omega$). The angled blue ellipse shows the orbital plane, with the periapsis point marked. The grey plane is parallel to the vector pointing toward the observer and gives perspective for how the orbital plane is inclined with respect to Earth.}
\label{fig:angles}
\end{figure}

The relative right ascension and declination values ($\Delta \alpha$ and $\Delta \delta$) with respect to the celestial coordinates of the \BB\ in the sky (denoted by $\alpha_0$ and $\delta_0$) use combinations of these constants \citep[as described in][]{2019AJ....158....4O}, such that
\begin{equation}\label{eq:deltaRA}
   \Delta\alpha = \alpha - \alpha_0 = \frac{1}{d} (Bx + Gy)\;, 
\end{equation}
and
\begin{equation}\label{eq:deltaDec}
    \Delta\delta = \delta - \delta_0 = \frac{1}{d} (Ax + Fy)\;,
\end{equation}
where $d$ is the distance between the \SSB\ and the \BB. These sky projection values are in units of radians.

While proper motion and parallax do fall under the category of astrometry, since they affect the observed positions of the binary system in the sky, they are calibrated out by subtracting the \BB\ coordinates from the observed right ascension and declination values, as done in Equations~(\ref{eq:deltaRA}) and (\ref{eq:deltaDec}). The \BB\ coordinates, $\alpha_0$ and $\delta_0$, change over time due to proper motion. Given some sky coordinate, $a_i$ and $\delta_i$, at initial time $t_0$, after time $t_\text{obs}$, the new right ascension becomes
\begin{equation}\label{eq:alpha0}
    \alpha_0(t) = \alpha_i + \mu_\alpha (t_\text{obs}-t_0)
\end{equation}
and the new declination becomes
\begin{equation}\label{eq:delta0}
    \delta_0(t) = \delta_i + \mu_\delta (t_\text{obs}-t_0)\;.
\end{equation}

Parallax and proper motions still contribute to line-of-sight motions, however, and we discuss the effects on observed radial velocities in Section~\ref{subsec:KB}.
\end{subequations}

The various Newtonian and relativistic effects enter the astrometric model in two ways, via their contribution to: {\em (i)\/} the relative positions of the stars and the black hole in the frame of the binary barycenter and {\em (ii)\/} the light propagation time between the star and the observer. Indeed, Equation~(\ref{eq:kepler}) for the calculation of the relative positions is written in terms of $t_{\rm star}$, which is determined by propagation effects and is related to the time of observation $t_{\rm obs}$ via Equation~(\ref{eq:tobs-tstar}). We provide explicit equations for the various propagation effects in the following section, since we will use the same equations to derive the various spectroscopic effects. 

In the astrometric model, so far, we have neglected the effects of gravitational lensing of the stellar positions in the black-hole spacetime. These effects have been explored, e.g., in \citep{2004MNRAS.355L...6N} and \citet{2012ApJ...753...56B}, and, albeit non-detectable with current data, might be relevant for modeling future observations \citep{2017A&A...608A..60G}.

\section{Spectroscopic Model}\label{sec:spec-model}

As discussed in Section~\ref{sec:intro}, pulsar timing models (e.g., \citetalias{DD2}; \citealt{Tempo2006}) use the time delays between emission and observation of pulses from binary pulsar systems to fit for the orbital parameters. With the S stars, we have both astrometric as well as spectroscopic data. In the previous section, we presented the equations we use to model the astrometric data. In this section, we present new line-of-sight velocity equations, which we derive from the time delay equations given in \citet{Tempo2006}.

Our spectroscopic model incorporates five timing effects (presented in the following subsections): the binary Roemer effect (\S\ref{subsec:RB}), the periastron precession (\S\ref{subsec:periastron}), the binary Einstein effect (\S\ref{subsec:EB}), the binary Shapiro effect (\S\ref{subsec:SB}), and the Earth Roemer effect (\S\ref{subsec:RO}). We also derive velocity equations for the Kopeikin effect (\S\ref{subsec:KB}) and the solar Shapiro effect (which we do not report here for brevity) to check their magnitudes and confirm that they are negligible. This allows us to omit other timing delays listed in \citet{Tempo2006} that are caused by Earth, the solar system, interstellar space, as well as higher-order GR effects, such as frame dragging, quadrupole moment, and additional 2PN terms described in \citet{2010ApJ...720.1303A}.

All time derivatives listed in this section are with respect to the observer time ($t_\text{obs}$), or the arrival time of the light from the star as seen by the observer.

\subsection{Binary Roemer Effect}\label{subsec:RB}
The binary Roemer delay component (i.e., the component of the Roemer delay that pertains only to the movement of the star around the \BB) is denoted by \RB. It is simply the light travel time for the line-of-sight distance across the orbit divided by the speed of light (i.e., $ \RB = -\hat{n} \cdot \vec{r}/c$). Adapting notation from both \citetalias{DD2} and \citet{Tempo2006}, we calculate the binary Roemer delay as: 
\begin{subequations}
\begin{align}\label{eq:RB}
    \RB = -a_r [&\left(\cos u - e_r\right) \sin \omega \sin i \nonumber\\
    + & \sin u \left(1- e_{\theta}^2 \right)^{1/2} \cos \omega \sin i ]\;.
\end{align}

Taking the time derivative of Equation~(\ref{eq:RB}) gives the equation of radial velocity resulting from the binary Roemer effect
\begin{align}\label{eq:dRBdt}
    \frac{d\Delta_\text{RB}}{dt}=-a_r\frac{du}{dt} \sin i \Bigl[&\cos u \cos \omega \left(1-e_\theta^2\right)^{1/2}\nonumber\\
    - &\sin u \sin \omega \Bigr] + v_\omega\;,
\end{align}
where $du/dt$ is derived from the Kepler Equation (Equation~\ref{eq:kepler}) such that
\begin{equation}\label{eq:dudt}
    \frac{du}{dt}=\frac{2\pi}{P}\frac{1}{1-e_t \cos u}\;.
\end{equation}
\end{subequations}
The additional term $v_\omega$ results from the time dependence of the argument of periastron, $\omega$. It is known as the \textit{periastron precession}, which we introduce below.

\subsection{Periastron Precession}\label{subsec:periastron}
Periastron precession is a first-order general relativistic effect that causes the orbit of an object to shift or rotate around its pericenter over time. As the location of periapsis advances around the \BB, the slight change in direction from a regular elliptical orbit results in an observed, line-of-sight velocity boost for the orbiting object. The $d\omega/dt$ term in the time derivative of the binary Roemer delay (Equation~\ref{eq:dRBdt}) describes this spectroscopic effect,
\begin{subequations}
\begin{align}\label{eq:vomega}
    v_\omega = -a_r\frac{d\omega}{dt} \sin i \Bigl[&\cos u \cos \omega \nonumber\\
    -& \sin u \sin \omega \left(1-e_\theta^2\right)^{1/2}\Bigr]\;.
\end{align}
The time derivative of the argument of periastron ($d\omega/dt$) relates to the time derivative of mean anomaly $du/dt$ (Equation~\ref{eq:dudt}) by
\begin{align}\label{eq:domega-dt}
\frac{d\omega}{dt} &=(K-1)\frac{dA_e(u)}{dt}\nonumber\\
&=(K-1)\frac{\left(1-e_\theta^2\right)^{1/2}}{1-e_\theta \cos u} \frac{du}{dt}\;,
\end{align}
where $K$ is again the dimensionless parameter related to the total momentum of the system and determines the amount of orbital precession (Equations~\ref{eq:K}-\ref{eq:precess} and \ref{eq:omega_tdep}).
\end{subequations}

\subsection{Binary Einstein Effect}\label{subsec:EB}
The binary Einstein delay (\EB) accounts for the difference between the star time of emission ($t_\text{star}$) and the \BB\ time (\tBB), which occurs due to both gravitational redshift (from GR) and time dilation (from special relativity).
We write the binary Einstein time delay as
\begin{subequations}
\begin{equation}\label{eq:EB}
    \EB = \frac{e_t}{n}\frac{m_1 (q+2)}{a_R(1+q)} \sin u\;,
\end{equation}
where $m_1$ is the companion mass to the orbiting star (since we define $m_1\geq m_2$). The derivative of this time delay, then, is simply
\begin{equation}
   \frac{d\Delta_\text{EB}}{dt} = \frac{e_t}{n}\frac{m_1 (q+2)}{a_R(1+q)} \cos u \frac{du}{dt}\;,
\end{equation}
which, when we substitute in the expression for $du/dt$ (Equation~\ref{eq:dudt}), becomes
\begin{equation}\label{eq:dEBdt}
   \frac{d\Delta_\text{EB}}{dt} = \frac{e_t}{1-e_t \cos u}\frac{m_1 (q+2)}{a_R(1+q)} \cos u\;.
\end{equation}

In the case of a circular orbit, Equation~(\ref{eq:dEBdt}) results in a null effect, even though gravitational and relativistic Doppler effects are still present. The reason is that this equation only describes the {\em change} in the combined wavelength shift along an eccentric orbit introduced by the changing separation and velocity magnitude. The baseline effect is absorbed into the systemic radial velocity of the object, i.e., the line-of-sight proper motion ($\mu_\parallel$), which is a free parameter, and can be trivially obtained in post-processing.

\end{subequations}

\subsection{Binary Shapiro Effect}\label{subsec:SB}
The binary Shapiro delay (\SB) describes the general relativistic effect introduced when light from one of the bodies in the model passes through the gravitational well of the other. We rewrite Equation~(73) from \citet{Tempo2006} in geometrized units as
\begin{subequations}
\begin{align}\label{eq:SB}
    \SB = -2 m_1 \ln \Bigl\{&1-e\cos u-\sin i\bigl[\sin\omega(\cos u - e) \nonumber \\
    &+ \sqrt{1-e^2}\cos\omega\sin u\bigr]\Bigr\}\;,
\end{align}
where the usage of $e$ without a subscript indicates that any eccentricity (i.e., $e_R$, $e_\theta$, $e_t$) may be used, as the difference is negligible. This results from the fact that the expressions for the three eccentricities differ at the $\sim1/c^2$ order, which we omit in the 1PN limit since they already multiply a term of the same order.

The time derivative of the binary Shapiro delay (\SB)\ gives the binary Shapiro effect for spectroscopic measurements:
\begin{align}\label{eq:dSBdt}
    \frac{d\Delta_\text{SB}}{dt}=-&\frac{2 m_1}{\mathcal A} \frac{du}{dt}\Bigl[e \sin u - \sin i \nonumber\\
    &\Bigl(\cos \omega \cos u \sqrt{1-e^2} - \sin \omega \sin u \Bigr) \Bigr]\;,
\end{align}
where we have introduced the symbol $\mathcal{A}$, i.e.,
\begin{align}\label{eq:SB-A}
    \mathcal{A} \equiv 1 - e \cos u - \sin i \bigl[&\sin \omega \left(\cos u - e\right)\nonumber\\
    &+ \cos \omega \sin u \sqrt{1-e^2}\bigr]\;.
\end{align}

\end{subequations}

\subsection{Kopeikin Effect}\label{subsec:KB}
\begin{subequations}
The Kopeikin effect combines spectroscopic contributions from both proper motion of the binary system and the parallax as seen from Earth. As discussed in Section~\ref{sec:astromet}, this is an astrometric effect, as well, but is calibrated out by subtracting the \BB\ location from all observed positions of the two bodies for each time step or observation.

We base our definition of the Kopeikin effect (\KB) on Sections 2.7.1 and 2.7.2 in \citet{Tempo2006}, where it is broken into three components \citep[][their Eq.~72]{Tempo2006}:
\begin{equation}\label{eq:KB}
    \KB = \SR + \AOP + \OP\;.
\end{equation}
Here, \SR\ is caused by changes in the viewing angle geometry due to the proper motion, \AOP\ is due to the annual orbital parallax, and \OP \ is due to the orbital parallax (the orbital equivalent of the Shklovskii effect). These three components, rewritten in geometrized units from Edwards et al. (\citeyear{Tempo2006}, their Eqs.~73-75), are
\begin{align}\label{eq:SR}
    \SR =  a_r \sin i &\left(\taBB - t_0\right) \times\nonumber\\ \Bigl[ &(\mu_{\alpha*} \sin \Omega + \mu_\delta \cos \Omega) ~C \csc i \nonumber\\
    &+ (\mu_{\alpha*} \cos \Omega - \mu_\delta \sin \Omega) ~S \cot i\Bigr]\;,
\end{align}
\begin{align}\label{eq:AOP}
    \AOP = - \frac{a_R}{d_\text{AOP}} &\sin i \times\nonumber\\ \Bigl[&(\vec{r}_\earth \cdot \hat{\alpha} \sin \Omega + \vec{r}_\earth \cdot \hat{\delta} \cos \Omega) ~C \csc i \nonumber\\
    &+ (\vec{r}_\earth \cdot \hat{\alpha} \cos \Omega - \vec{r}_\earth \cdot \hat{\delta} \sin \Omega) ~S \cot i\Bigr]\;,
\end{align}
and
\begin{equation}\label{eq:OP}
    \OP = \frac{a_R^2}{2d_\text{OP}} (C^2 \csc^2 i + S^2 \cot^2 i)\;,
\end{equation}
where $C$ and $S$ are given by 
(65) and (66) in \citet{Tempo2006} and $\mu_{\alpha*} = \mu_\alpha \cos \delta$.

The annual orbital parallax and orbital parallax distances (to the \BB) are approximately equal such that $d \approx d_\text{AOP} \approx d_\text{OP}$. As with the previous effects, the total Kopeikin effect is simply the time derivative of the above equations,
\begin{equation}\label{eq:dKBdt}
    \frac{d\KB}{dt} = \frac{d\SR}{dt} + \frac{d\AOP}{dt} + \frac{d\OP}{dt}\;,
\end{equation}
where:
\begin{align}\label{eq:dSRdt}
    \frac{d\SR}{dt} = a_r &\sin i \left(\taBB - t_0\right)\nonumber\\ \Bigl[&\left(\mu_{\alpha*} \sin \Omega + \mu_\delta \cos \Omega\right) \frac{dC}{dt} \csc i\nonumber\\
    &+ \left(\mu_{\alpha*} \cos \Omega - \mu_\delta \sin \Omega\right)\frac{dS}{dt} \cot i\Bigr]\nonumber\\
    &+ \frac{\SR}{\taBB - t_0}\;,
\end{align}
\begin{equation}\label{eq:dAOPdt}
    \frac{d\AOP}{dt} = -\frac{a_r}{d_\text{AOP}} \sin i \left(\mathcal{B} + \mathcal{C}\right)\;,
\end{equation}
and
\begin{equation}\label{eq:dOPdt}
    \frac{d\OP}{dt} = \frac{a_r^2}{d_\text{OP}} \sin^2 i\left(C \frac{dC}{dt}\csc^2 i + S\frac{dS}{dt}\cot^2 i\right)\;.
\end{equation}
Here, the symbols $\mathcal{B}$ and $\mathcal{C}$ are defined as:
\begin{align}
    \mathcal{B} \equiv &\left(\vec{r}_\oplus \cdot \hat{\alpha} \sin \Omega + \vec{r}_\oplus \cdot \hat{\delta} \cos \Omega\right) \frac{dC}{dt} \csc i\nonumber\\
    + &\left(\vec{r}_\oplus \cdot \hat{\alpha} \cos \Omega - \vec{r}_\oplus \cdot \hat{\delta}\sin\Omega\right) \frac{dS}{dt}\cot i
\end{align}
and 
\begin{align}\label{eq:KB-C}
    \mathcal{C} \equiv &\frac{d\left(\vec{r}_\oplus \cdot \hat{\alpha}\right)}{dt} \left(\sin \Omega ~ C \csc i + \cos \Omega ~ S \cot i\right)\nonumber\\
    + &\frac{d\left(\vec{r}_\oplus \cdot \hat{\delta}\right)}{dt} \left(\cos \Omega ~ C \csc i + \sin \Omega ~ S \cot i\right)\;.
\end{align}

The time derivatives of the expressions for $C$ and $S$ are
\begin{align}\label{eq:dCdt}
    \frac{dC}{dt}  = -&\frac{du}{dt} \left[\cos \omega \sin u + \sin \omega \cos u \sqrt{1-e_\theta^2}\right]\nonumber\\
    - &\frac{d\omega}{dt} \left[\sin\omega (\cos u - e_r) - \cos \omega \sin u \sqrt{1-e_\theta^2}\right]
\end{align}
and
\begin{align}\label{eq:dSdt}
    \frac{dS}{dt}  = -&\frac{du}{dt} \left[\sin \omega \sin u + \cos \omega \cos u \sqrt{1-e_\theta^2}\right]\nonumber\\
    + &\frac{d\omega}{dt} \left[\cos\omega (\cos u - e_r) - \sin \omega \sin u \sqrt{1-e_\theta^2}\right]\;.
\end{align}
\end{subequations}

\subsection{Earth Roemer Effect}\label{subsec:RO}
The Earth Roemer effect is analogous to the binary Roemer effect (\S~\ref{subsec:RB}), except that the changes in light travel time are due to the motion of the Earth around the \SSB. We define the Earth-motion Roemer delay component using similar notation to Equations~(13)-(16) in \citet{Tempo2006}:
\begin{subequations}
\begin{equation}\label{eq:RO}
    \RO = -\vec{r}_\earth \cdot \hat{R}_\text{BB}\;,
\end{equation}
where $\vec{r}_\earth$ is the vector from the \SSB \ to the Earth:
\begin{equation}\label{eq:rEarth}
    \vec{r}_\earth = \begin{pmatrix}x_\earth\\ y_\earth\\z_\earth\end{pmatrix}\;,
\end{equation} 
and $\hat{R}_\text{BB}$ is the unit vector between the \BB \ and the observer:
\begin{align}\label{eq:RBB}
    \hat{R}_\text{BB} = &\hat{\eta} + \mu_{\perp}\left(\taBB-t_0\right) \nonumber\\
    &- \left(\frac{1}{2}|\mu_{\perp}|^2 \hat{\eta} + \mu_{\perp}\mu_{\parallel}\right)\left(\taBB-t_0\right)^2\;.
\end{align}

The \BB-observer unit vector ($\hat{R}_\text{BB}$) breaks down to the primary component ($\hat{\eta}$), which is the initial unit vector between the observer and the \BB\ (in celestial coordinates), i.e.,
\begin{equation}\label{eq:etaHat}
   \hat{\eta} = \begin{pmatrix} \cos\alpha \cos\delta \\ \sin\alpha \cos\delta \\ \sin\delta\end{pmatrix} \;,
\end{equation}
and the shift in the \BB\ position over time due to proper motion ($\mu_\alpha$, $\mu_\delta$, $\mu_\parallel$) of the binary system. While the line-of-sight proper motion $\mu_\parallel$ is one of the free parameters for the model, the transverse proper motion $\mu_\perp$ depends on the right ascension and declination proper motions, $\mu_\alpha$ and $\mu_\delta$. We define the transverse proper motion as
\begin{equation}\label{eq:muPerp}
    \mu_{\perp} = \mu_{\alpha*}\hat{\alpha} + \mu_{\delta}\hat{\delta}\;,
\end{equation}
where the projected right ascension and declination vectors $\hat{\alpha}$ and $\hat{\delta}$ are given by Edwards et al. (\citeyear{Tempo2006}, their Eqs.~17-18),
\begin{equation}\label{eq:alphaHat}
    \hat{\alpha} = \begin{pmatrix} -\sin\alpha \\ \cos\alpha \\ 0 \end{pmatrix}\;,
\end{equation}
and
\begin{equation}\label{eq:deltaHat}
    \hat{\delta} = \begin{pmatrix} -\cos\alpha \sin\delta \\ -\sin\alpha \sin\delta \\ \cos\delta \end{pmatrix}\;.
\end{equation}

The Earth Roemer effect is the time derivative of the Earth Roemer delay (Equation~\ref{eq:RO}), such that
\begin{equation}\label{eq:dROdt}
    \frac{d\RO}{dt} = -\left(\mathcal{D}_1 + \mathcal{D}_2 + \mathcal{D}_3 + \mathcal{D}_4 \right)\;,
\end{equation}
where the symbols $\mathcal{D}_1$, $\mathcal{D}_2$, $\mathcal{D}_3$, and $\mathcal{D}_4$ are shorthand for
\begin{equation}\label{eq:D1}
    \mathcal{D}_1 = -\left(\vec{r}_\oplus \cdot \hat{\eta}\right) \left[\lvert \vec{\mu}_\perp \rvert^2 \left(\taBB - t_0\right)\right]\;,
\end{equation}
\begin{equation}\label{eq:D2}
    \mathcal{D}_2 = \frac{d\left(\vec{r}_\oplus \cdot \vec{\eta}\right)}{dt} \left[1 - \frac{1}{2} \lvert \vec{\mu}_\perp \rvert^2 \left(\taBB - t_0\right)^2\right]\;,
\end{equation}
\begin{equation}\label{eq:D3}
    \mathcal{D}_3 = \left(\vec{r}_\oplus \cdot \hat{\eta}\right) \left[1- \mu_\parallel \left(\taBB - t_0\right) \right]\;,
\end{equation}
and 
\begin{equation}\label{eq:D4}
    \mathcal{D}_4 = \frac{d\left(\vec{r}_\oplus \cdot \vec{\mu}_\perp\right)}{dt} \left[1 - \frac{1}{2} \lvert \vec{\mu}_\perp \rvert^2 \left(\taBB - t_0 \right)^2\right]\;.
\end{equation}
\end{subequations}

\section{Summary of Model}\label{sec:orbital-model-summary}
Having presented all of the equations and parameters we use in our analytic, 1PN model for S stars, we summarize in Figure~\ref{fig:algorithm} the steps one should take to implement it.

\begin{figure*}[htb]
    \centering
    \begin{tabular}{|p{2\columnwidth}|}
        \hline
        \textbf{Observed parameters:}\\
        Sky position of orbiting star ($\alpha(t)$ and $\delta(t)$), line-of-sight velocity ($v_z(t)$)\\
        \textbf{Free parameters:} \\
        Period ($P$), total mass ($M$), mass ratio ($q$), eccentricity ($e_R$), distance to system ($d$), orientation of system ($i$, $\Omega$, $\omega_0$), sky position of \BB\ ($\alpha_0(t)$, $\alpha_i$, $\delta_0(t)$, $\delta_i$), \BB\ proper motion ($\mu_\alpha$, $\mu_\delta$, $\mu_\parallel$), initial time of periapsis (or epoch of position, $t_0$)\\
        \hline
        \textbf{Steps:}
        \begin{enumerate}
            \item Using the input parameters, calculate:
            \begin{enumerate}
                \item Individual parameters $a_{r_1}$, $e_{r_1}$, $a_{r_2}$, and $e_{r_2}$ with Equations~(\ref{eq:ar1})-(\ref{eq:er2})
                \item General eccentricities $e_t$ and $e_\theta$ with Equations~(\ref{eq:et}) and (\ref{eq:etheta})
                \item Dimensionless reduced mass $\nu$ with Equation~(\ref{eq:nu})
                \item Total energy $E$, total momentum $J$, and $K$ parameter with Equations~(\ref{eq:E})-(\ref{eq:K})
            \end{enumerate}
            \item For each time step:
            \begin{enumerate}
                \item Solve for the mean anomaly $u$ using the Newton-Raphson method (or similar fast algorithm) for Equation~(\ref{eq:kepler}).
                \item Calculate the object distances ($r_1$, $r_2$) from the \BB\ with Equations~(\ref{eq:r1}) and (\ref{eq:r2}), the angle $\theta$ via Equation~(\ref{eq:theta}), and the time-dependent argument of periapsis $\omega(u)$ with Equation~(\ref{eq:omega_tdep}).
                \item Convert the polar, binary barycentric orbits ($r$/$\theta$) to Cartesian coordinates ($x$/$y$/$z$) with Equations~(\ref{eq:r1-to-x1})-(\ref{eq:r2-to-y2}).
                \item Project the Cartesian, binary barycentric orbits ($x$/$y$/$z$) to the sky plane ($\Delta \alpha$/$\Delta \delta$) to obtain astrometric data relative to the \BB\ location ($\alpha_0$/$\alpha$) with Equations~(\ref{eq:TH-A})-(\ref{eq:delta0}).
                \item Calculate the spectroscopic effects
                \begin{enumerate}
                    \item Binary Roemer effect with Equations~(\ref{eq:dRBdt})-(\ref{eq:dudt})
                    \item Periastron Precession with Equations~(\ref{eq:vomega})-(\ref{eq:domega-dt}) and (\ref{eq:dudt})
                    \item Binary Einstein effect with Equation~(\ref{eq:dEBdt})
                    \item Binary Shapiro effect with Equations~(\ref{eq:dSBdt})-(\ref{eq:SB-A}) and (\ref{eq:dudt})
                    \item Kopeikin effect with Equations~(\ref{eq:dKBdt})-(\ref{eq:dSdt}), (\ref{eq:SR}), (\ref{eq:dudt}), and (\ref{eq:domega-dt})$^\dagger$
                    \item Earth Roemer effect with Equations~(\ref{eq:dROdt})-(\ref{eq:D4}), (\ref{eq:etaHat})-(\ref{eq:muPerp}), and (\ref{eq:rEarth})$^\dagger$
                \end{enumerate}
            \end{enumerate}
        \end{enumerate}
        \footnotesize{$^\dagger$ Calculations for these spectroscopic effects require ephemerides for the Earth, Sun, and Sgr~A* in ICRS coordinates.}\\
        \hline
    \end{tabular}
    \caption{Steps for implementing the new 1PN astrometric and spectroscopic model presented in this paper.}
    \label{fig:algorithm}
\end{figure*}

The Earth Roemer and Kopeikin effects both depend on the position of the Earth around the Sun, and so to calculate them, we must use Earth ephemerides. For the time derivatives of any values that use these data, we use three-point midpoint differentiation.

\section{Spectroscopic Effects of S stars}\label{sec:spec-effecs}

\subsection{Magnitude of Spectroscopic Effects}\label{sec:mags}

Due to the exquisite accuracy of pulsar observations, pulsar timing models must incorporate numerous time delays in addition to the geometric and relativistic effects from the binary system. Such effects include atmospheric delays, dispersion from both the interplanetary medium and interstellar medium, frequency-dependent delays, and gravitational effects from many bodies in our solar system, namely Venus, Jupiter, Saturn, Uranus, and Neptune (in addition to the Sun).

Timing models for spectroscopic observations of S stars, on the other hand, do not require the same accuracy. The current sensitivity level for Doppler shifts on the instruments capable of observing the S stars (i.e., Keck and the Very Large Telescope) are around 10~\kms\ \citep{1998SPIE.3353..704T,2018SPIE10702E..0AM} and upcoming ground-based 25-40~m extremely large telescopes (ELTs) are anticipated to have velocity sensitivities of around 1~\kms\ \citep[e.g.,][]{2019BAAS...51g.134M,2021Msngr.182...27M}. As a result, we may neglect timing effects that are significantly smaller than predicted ELT-class sensitivities. 
Note that the ELT-class sensitivities have been estimated for stars down to $\sim 19$ mid-infrared magnitude, but will depend on the actual properties of stars with smaller orbital separations that may be discovered in the future.

In considering the possibility of detecting higher-order general relativistic effects, it is useful to consider the order-of-magnitude strengths of the spectroscopic effects described in Section~\ref{sec:spec-model}. We derived scaling equations for the six velocity components of the spectroscopic model and present them in Appendix~\ref{appendix:mag-eqs}.

\begin{figure*}[htb!]
\centering
\includegraphics[trim=1.5cm 0.5cm 0.5cm 0.5cm, clip, width=\textwidth]{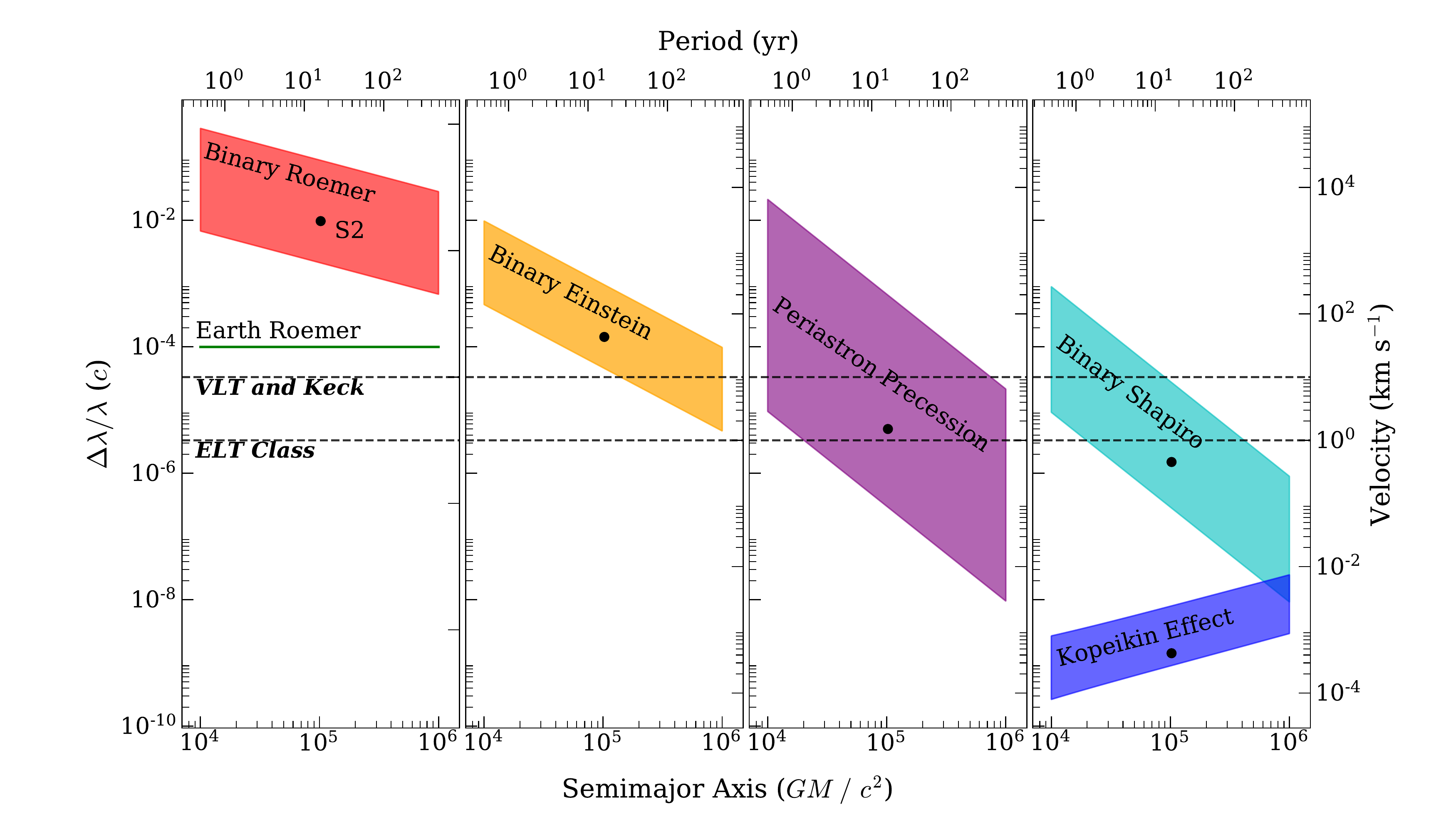}
\caption{Absolute magnitudes of characteristic contributions to the radial-velocity corrections in spectroscopic models of stellar orbits around Sgr~A* introduced by the various Newtonian and post-Newtonian effects. Each shaded area corresponds to orbital eccentricities in the range 0.7 - 0.98, with lower limits corresponding to $e=0.7$ and upper limits corresponding to $e=0.98$. The orientation angles of the orbits are assumed to be those of the S0-2 star. Horizontal dashed lines show the current measurement limits for VLT/Keck observations as well as the expected limits for an ELT-class telescope. We see that while the binary Roemer effect (far left, solid red band) is the dominant velocity contribution, the three GR effects -- binary Einstein effect, periastron precession, and binary Shapiro effect -- rapidly increase at smaller semimajor axes/orbital periods. The periastron precession effect (second from right, purple band) has the strongest dependence on eccentricity and spans four orders of magnitude, overlapping with the binary Einstein effect (second from left, yellow band) and Shapiro effect (far right, light blue band), which span one and three orders of magnitude, respectively. While the binary system spectroscopic effects -- binary Roemer, binary Einstein, periastron precession, and binary Shapiro -- decrease for larger orbital separations (semimajor axes), the Kopeikin effect describes velocity contributions due to parallax and proper motions that are due to the movement of the observer and, as such, increases for larger orbital separations.}
\label{fig:mag-velocities}
\end{figure*}

Using these scaling relations, we show in Figure~\ref{fig:mag-velocities} the relative strengths of the binary Roemer effect, binary Einstein effect, periastron precession, binary Shapiro effect, Earth Roemer effect, and Kopeikin effect for orbital periods that range from 0.5~years to 500~years with a fixed system mass and two-body mass ratio that is characteristic of Sgr~A* and the S stars. The widths of the bands result from a range of  eccentricities, $e=0.7-0.98$, with the smaller effects corresponding to lower eccentricities.

In this plot, we see that the absolute magnitudes of the spectroscopic effects differ greatly from each other: while the binary Roemer effect results in velocities that range from $10^2 - 10^5$~\kms, the Kopeikin effect has velocities that span $10^{-4}-10^{-2}$~\kms. The binary Einstein, periastron precession, and binary Shapiro effects fall between these two extremes, with Doppler effect ranges of $10^0-10^3$~\kms, $10^{-3}-10^4$~\kms, and $10^{-3}-10^2$~\kms, respectively.

Another notable aspect of Figure~\ref{fig:mag-velocities} is the difference in the slopes of the various spectroscopic effects. As a purely geometric phenomenon, wavelength shifts introduced by the binary Roemer effect scale as $\Delta \lambda/\lambda \sim a_R^{-1/2} \sim P^{-1/3}$, growing larger with smaller semimajor axes. The magnitudes of the periastron precession and binary Shapiro effects also increase with smaller semimajor axes and periods, although they do so much more rapidly ($\Delta \lambda/\lambda \sim a_R^{-3/2} \sim P^{-1}$). This is because of the steeper scaling of the PN corrections with orbital separation. Similarly, the binary Einstein effect lies in between these scalings ($\Delta \lambda/\lambda \sim a_R^{-1} \sim P^{-2/3}$). The Kopeikin effect, on the other hand, is not a GR effect. Rather, it describes the radial velocity contributions from the parallax and proper motions of the binary system. As a result, since there is more time for parallactic effects and proper motions to affect observations for any given orbit, the magnitude of the corresponding Doppler effect grows as the orbital period increases ($\Delta \lambda/\lambda \sim a_R \sim P^{2/3}$). The magnitude of the Earth Roemer spectroscopic effect is independent of binary orbital period and, as such, has a constant magnitude of $\sim 30$~\kms. 

One last property to note is the effect of eccentricity on the magnitudes of the spectroscopic effects. For any fixed orbital period, the binary Roemer effect has a span of two orders of magnitude, the binary Einstein effect ranges around two orders, the periastron precession stretches almost four orders of magnitude, the binary Shapiro effect extends over three orders, and the Kopeikin effect spans one order. The three GR effects -- binary Einstein effect, periastron precession, and binary Shapiro effect -- are most dramatically influenced by changes in eccentricity. Higher eccentricities bring orbiting stars increasingly closer to the central black hole, resulting in higher angular momenta of the system (Equation~\ref{eq:J}-\ref{eq:K}) and larger amounts of orbital precession ($\Delta\theta$), as well as greater Doppler contributions from the periastron precession effect. Similarly, the smaller the periapsis distance is (due to higher eccentricities), the closer the light from orbiting stars must pass through the potential well of the central black hole, which also increases the binary Shapiro effect.

We can reach several significant conclusions from Figure~\ref{fig:mag-velocities}. First, based on the expected sensitivities of ELT-class telescopes of $\sim$1~\kms, we find that the binary Shapiro effect should be detectable by these telescopes through spectroscopic observations alone during the next S0-2 periapsis passage in 2034. Second, if observations detect and confirm S stars with shorter periods ($\lesssim10$~yr) and high eccentricities ($\gtrsim0.85$ for periods of less than a year and $\gtrsim0.95$ for periods of less than 10~yr), existing telescopes with velocity sensitivities of $\sim$10~\kms\ may already be capable of detecting the first-ever binary Shapiro effect for any S star.

\subsection{Fingerprints of Spectroscopic Effects}
\begin{figure*}[htb!]
\centering
\includegraphics[width=0.7\textwidth]{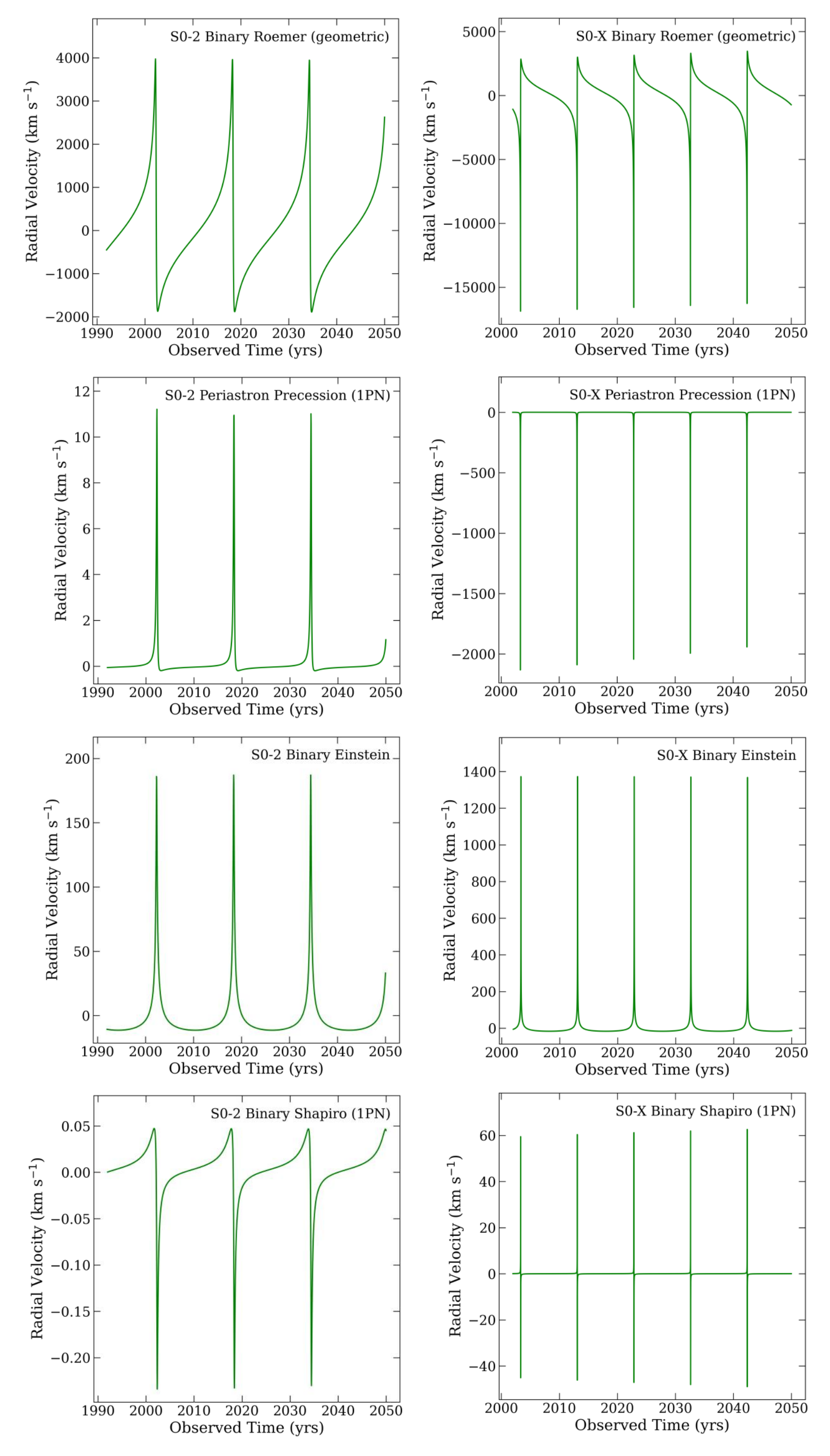}
\caption{The contributions of different Newtonian and post-Newtonian effects to the line-of-sight velocities of (left) the S0-2 star and (right) S0-X, a hypothetical, sub-10 yr period star, around Sgr~A*. Unique functional forms for the different spectroscopic effects make it easy to disentangle geometric, Earth-related, and solar-system-related effects from true GR phenomena. The eccentricity and orientation of its orbit make post-Newtonian effects in a star like S0-X potentially detectable even with current instruments.}
\label{fig:s2-s62-grid}
\end{figure*}

One of the key advantages of the spectroscopic model we presented in Section~\ref{sec:spec-model} is that each effect has a unique signature (or ``fingerprint") that we can search for in the data. Figure~\ref{fig:s2-s62-grid} shows the four binary system spectroscopic effects for the S0-2 star and a hypothetical star with period less than ten years, which we refer to as S0-X. For the sake of illustration, we picked a set of fiducial parameters similar to those listed in \citet{2020ApJ...889...61P}, although such a star has yet to be confirmed \citep{2022A&A...657A..82G}. Table~\ref{table:s02-s0x-vals} in Appendix~\ref{appendix:s02-s0x-params} lists the parameter values used for the S0-2 and S0-X models.

For S0-2, the binary Roemer effect (left, top) stretches from $-2000$~\kms\ to $+4000$~\kms, whereas the periastron precession (left, second from top), binary Einstein (left, second from bottom), and binary Shapiro effects (left, bottom) have ranges of $\sim 0$~\kms\ to $+11$~\kms, $0$ to $200$~\kms, and $-0.25$ to $+0.05$~\kms, respectively. For our hypothetical, sub-10 yr period star, S0-X, we see very different behaviors, with the binary Roemer effect (right, top) spanning $-15000$~\kms\ to $+3000$~\kms, the periastron precession (right, second from top) ranging $-2000$~\kms\ to $\sim0$~\kms, the binary Einstein effect (right, second from bottom) stretching from $0$ to $1400$~\kms, and the binary Shapiro effect (right, bottom) extending from $-40$~\kms\ to $+60$~\kms.

Clearly, the patterns of the three effects vary greatly, both with respect to each other and between stars. One of the primary factors driving this is the orientation of the stellar orbit with respect to the observer. Both the binary Roemer effect and the periastron precession relate to the geometry of the orbit (Newtonian and GR): the binary Roemer effect is the change in light-travel time and the precession of the periapsis point due to GR results in an additional line-of-sight velocity boost as the orbit precesses. As a result, both introduce radial velocity contributions that have the same signs. For S0-2, the effects are primarily positive, while for S0-X, they are primarily negative. 

The binary Shapiro effect, on the other hand, is stronger when emitted light from an orbiting star passes closer to the black hole en route to the observer. This is opposite of the effect that governs the sign of the binary Roemer and periastron precession effects, and so it primarily has the opposite sign of the other two. Depending on the orientation of the orbit with respect to the Earth, the binary Shapiro effect rapidly changes sign as the star moves behind the black hole.

While the three aforementioned spectroscopic effects all exhibit sign changes, the binary Einstein effect is always positive. This is because, by the nature of the gravitational redshift, the emitted light can only be redshifted.

The net effect of the different behaviors and signs of the spectroscopic effects is that they act like fingerprints on radial velocity observations over time. The spectroscopic signature of the periapsis precession (panels second from top) cannot be confused with that of the binary Shapiro effect (bottom panels), as they exhibit very different functional forms and signs. This is valuable in the analysis of S-star orbits, as it enables the identification of both binary system geometric and GR effects and solar system geometric and GR effects and also minimizes the chance of confusing them.

\begin{figure*}[htb!]
\centering
\includegraphics[trim=0.25cm 0.5cm 0cm 0cm, clip, width=2\columnwidth]{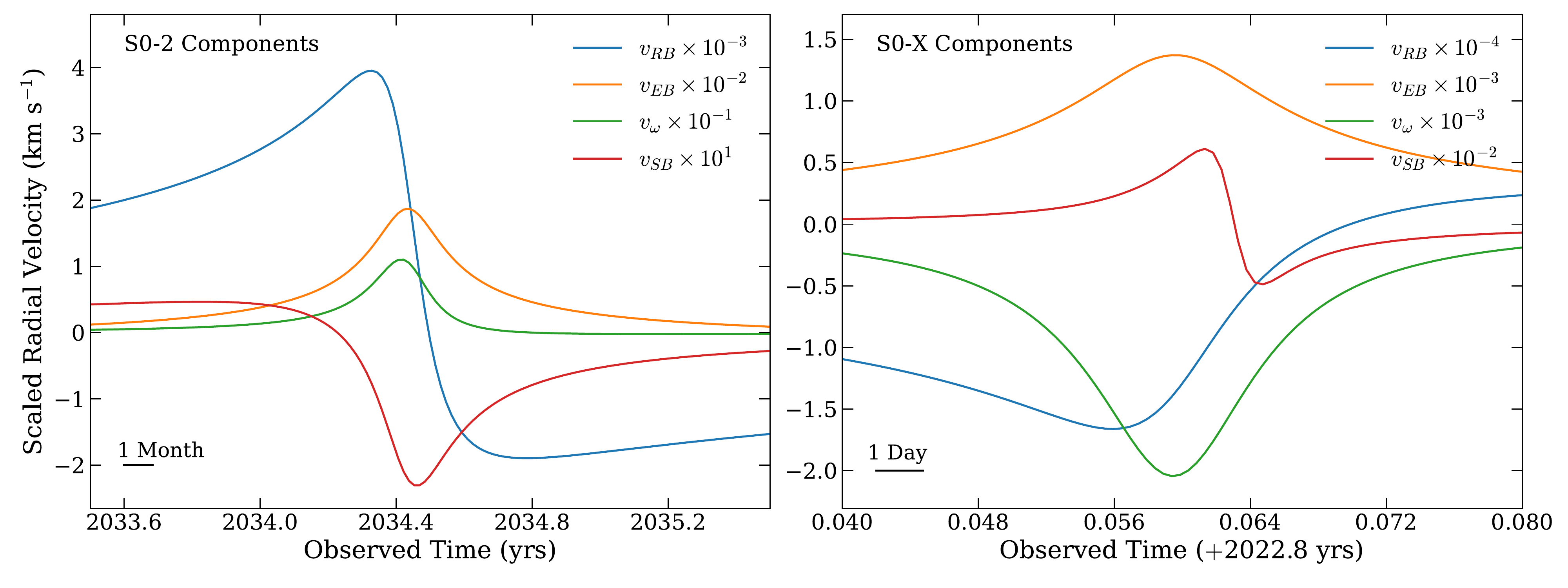}
\caption{Velocity contributions of the binary Roemer effect, binary Einstein effect, periastron precession, and binary Shapiro effect for the periapsis passages of S0-2 (left) and S0-X, a hypothetical, sub-10 year period star (right), around Sgr~A*. Velocities are scaled to better show the unique shapes of the velocity components over time. Timescale bars in the lower left of both plots show a one-month period and one-day period for S0-2 and S0-X periapsis passages, respectively, as an indicator of useful observational cadences. The peaks and troughs of the four spectroscopic effects are offset from each other in time. This implies that they will not cancel out, which means the residual-fitting method is a viable way to attempt to detect GR phenomena with this model.}
\label{fig:timing-velocities}
\end{figure*}

Figure~\ref{fig:timing-velocities} zooms in on the spectroscopic effects shown in Figure~\ref{fig:s2-s62-grid} to compare the timings of the maxima and minima of each of the radial velocity contribution effects. The velocities are scaled to similar orders of magnitude to elucidate these comparisons. We can see that all three effects are offset from each other. The binary Roemer effect peaks in magnitude first, the binary Einstein, second, the periastron precession, third, and the binary Shapiro, fourth. This is crucial, as it demonstrates that the velocity contributions will not cancel out and, therefore, fitting spectroscopic effects via the residual method is a viable way to detect them. \citet{2022A&A...660A..13H} have come to a similar conclusion for different effects in the astrometric domain.

\section{Conclusions}
The results from our astrometric and spectroscopic model have several key implications. First, calculations using the orbital parameters of the S0-2 star \citep{2020A&A...636L...5G} show that the binary Shapiro effect should be detectable spectroscopically by ELT-class telescopes during its next periapsis period. For any stars discovered with shorter periods and/or higher eccentricities, these effects could already potentially be detected with current instruments. Although stellar winds result in rotational broadening of lines, winds from B~stars like S0-2 (and other stars in the Galactic center) are weak and not directly measurable \citep{2008ApJ...672L.119M,2021PhRvD.103f3041F}. Furthermore, since observational studies of the S stars use broad absorption lines to determine line-of-sight velocities, the systematic uncertainties of $\sim 10$~\kms\ mean that detecting the described spectroscopic effects are still realistic. Overall, our model shows that while astrometric data are crucial for constraining the orbital parameters (i.e., period, orientation, eccentricity) of a particular star, they are not essential for detecting general relativistic effects, which can be done entirely via spectroscopy.

Second, the analytic nature of the model allows us to evaluate directly the various observables at any time, without having to integrate the differential equations of motion for each star. This results in a remarkable reduction of computational cost, as it can be easily demonstrated with fiducial values for the hypothetical S0-X star. For example, due to the S0-X orbital parameters and orientation, the velocity correction from the periapsis precession changes dramatically over the course of one-hundredth of its orbit by 1000 km~s\textsuperscript{-1}, i.e., by 100 times the magnitude of the current measurement uncertainties. Integrating the geodesic equations with a method of order $n$ (e.g., such that $n=4$ for a fourth order Runge-Kutta method) using a timestep $\Delta t$ would introduce a fractional truncation error in each timestep that is $\sim (\Delta t/10^{-2} P)^{n+1}$, where we have used the fact that there is significant evolution over a fraction $10^{-2}$ of the orbital period $P$. The total accumulated error after integrating for a single orbit will be $\sim (P/\Delta t)(\Delta t/10^{-2} P)^{n+1}$. Requiring for this error to be smaller than the measurement error, i.e., to be $\sim 10^{-3}$, leads to the conclusion that we would need at least $P/\Delta t\sim 100^{2+5/n}$ timesteps per orbit, which is $\gtrsim 10^4$ for $n\ge 2$. With our analytic model, evaluating the precise correction at any point in time will take only a handful of floating point operations. This factor of $\sim 10^4$ reduction in computational cost of using an analytic model compared to a numerical one implies that a Bayesian MCMC statistical study that takes an hour for the analytical model will need about a year for the numerical one. This allows more efficient parameter space exploration as well as the possibility of fitting multiple stellar orbits with simultaneous constraints. Upcoming ELT-class telescopes will likely be pivotal in producing more data for stars in the Galactic center over shorter integration times.

While the post-Newtonian approach has many computational and scientific advantages, it does not allow for a straightforward incorporation of the perturbing effects of an extended mass distribution~\citep[see, e.g.,][]{1985AmJPh..53..694J,2001A&A...374...95R}. However, the current limits on the mass of an unseen perturber are of the order of $\sim$3000~\Msun \citep{2020A&A...636L...5G,2022A&A...657L..12G,2022A&A...660A..13H} for extended mass within the S2 orbit and $\sim$1000~\Msun for a compact mass within the inner arcsecond of Sgr A* and within or around the orbit of S2 \citep{2010PhRvD..81f2002M,2020A&A...636L...5G}. Given the current accuracy of the astrometric and spectroscopic data, the presence of such a perturbing mass is not significant and does not compete with the effects that we have considered at the 1PN order.


\begin{acknowledgments}
We thank Norbert Wex and Gunther Witzel for carefully reading the manuscript and for their comments and suggestions. We also thank the members of the Extreme Astrophysics group at the University of Arizona for useful comments and suggestions throughout this project. This work was supported in part by NSF PIRE award OISE-1743747 and NSF award AST-1715061.
\end{acknowledgments}

\appendix

\section{Order of Magnitude Equations}
\label{appendix:mag-eqs}

To estimate the relative strengths of the various velocity effects, we simplified the equations in Section~\ref{sec:spec-model} to scaling relations, which are given below. These relations were used to calculate the values in Figure~\ref{fig:mag-velocities}.

\noindent (i) Binary Roemer Effect
\begin{align}
    \frac{d\RB}{dt} \approx &  \frac{1}{2} a^{-1/2} M^{1/2} \frac{\sin i}{1-e} \Bigl[\cos \omega \sqrt{1-e^2} - \sin \omega  \Bigr]\\
    \approx& -942 \left(\frac{a}{1000~\AU}\right)^{-1/2} \left(\frac{M}{4\times10^6~\Msun}\right)^{1/2} \frac{\sin i}{1-e} \Bigl[\cos \omega \sqrt{1-e^2} - \sin \omega  \Bigr]~\text{km s}^{-1}
\end{align}

\noindent (ii) Periastron Precession
\begin{align}
    v_\omega \approx& - a^{-3/2} M^{3/2} \frac{3}{\sqrt{1-e^2}} \frac{\sin i}{(1-e)^2} \frac{1}{2}\Bigl[\cos \omega \sqrt{1-e^2} - \sin \omega  \Bigr]\\
    \approx& -0.112 \left(\frac{a}{1000~\AU}\right)^{-3/2} \left(\frac{M}{4\times10^6~\Msun}\right)^{3/2} \frac{\sin i}{\sqrt{1-e^2}(1-e)^2} \Bigl[\cos \omega \sqrt{1-e^2} - \sin \omega  \Bigr]~\text{km s}^{-1}
\end{align}

\noindent (iii) Binary Einstein Effect
\begin{align}
    \frac{d\Delta_\text{EB}}{dt} &\approx m_1 a^{-1} \frac{e}{1-e} \frac{q+2}{q+1}\\
    &\approx 23.7 \left(\frac{m_1}{4\times10^{6}~\Msun}\right)\left(\frac{a}{1000~\AU}\right)^{-1} \frac{e}{1-e} \frac{q+2}{q+1}
\end{align}

\noindent (iv) Binary Shapiro Effect
\begin{align}
    \frac{d\Delta_\text{SB}}{dt} &\approx -2 a^{-3/2} M^{3/2} \frac{1}{1-e} \frac{e - \sin i \Bigl(\cos \omega \sqrt{1-e^2} - \sin \omega \Bigr)}{1 - e - \sin i \bigl[\sin \omega \left(1 - e\right) + \cos \omega \sqrt{1-e^2}\bigr]}\\
    &\approx -3.72\times10^{-2} \left(\frac{a}{1000~\AU}\right)^{-3/2} \left(\frac{M}{4\times10^{6}~\Msun}\right)^{3/2} \frac{1}{1-e} \frac{e - \sin i \Bigl(\cos \omega \sqrt{1-e^2} - \sin \omega \Bigr)}{1 - e - \sin i \bigl[\sin \omega \left(1 - e\right) + \cos \omega \sqrt{1-e^2}\bigr]}
\end{align}

\noindent (v) Kopeikin Effect
\begin{align}
    \frac{d \Delta_\text{SR}}{dt} \approx&  a \Bigl[\cos i \left(\sqrt{1-e^2} \cos \omega + (1-e) \sin \omega\right) \left(\mu_{\alpha*} \cos \Omega-\mu_\delta \sin \Omega\right)\nonumber\\
    &\qquad -\left(\sqrt{1-e^2} \sin \omega+(e-1) \cos \omega\right) \left(\mu_{\alpha*} \sin \Omega+\mu_\delta \cos \Omega\right)\Bigr]\nonumber\\
    &-a ^{-1/2}M^{1/2} t\frac{1}{1-e} \Bigl[\cos i \left(\sqrt{1-e^2} \cos \omega+\sin \omega\right) \left(\mu_{\alpha*} \cos \Omega-\mu_\delta \sin \Omega\right)\nonumber\\
    &\qquad- \left(\sqrt{1-e^2} \sin \omega+\cos \omega\right) \left(\mu_{\alpha*} \sin \Omega+\mu_\delta \cos \Omega\right)\Bigr]\\
    \approx& 2.30 \times10^{-5} \left(\frac{a}{1000~\AU}\right) \Bigl[\cos i \left(\sqrt{1-e^2} \cos \omega + (1-e) \sin \omega\right) \left(\mu_{\alpha*} \cos \Omega-\mu_\delta \sin \Omega\right)\nonumber\\
    &\qquad -\left(\sqrt{1-e^2} \sin \omega+(e-1) \cos \omega\right) \left(\mu_{\alpha*} \sin \Omega+\mu_\delta \cos \Omega\right)\Bigr]\nonumber\\
    &-4.57\times10^{-6} \left(\frac{a}{1000~\AU}\right)^{-1/2}\left(\frac{M}{4\times10^6 \Msun}\right)^{1/2} \left(\frac{t}{1~\text{yr}}\right)\frac{1}{1-e}\nonumber\\
    &\qquad \Bigl[\cos i \left(\sqrt{1-e^2} \cos \omega+\sin \omega\right) \left(\frac{\mu_{\alpha*}}{1~\text{mas yr}^{-1}} \cos \Omega-\frac{\mu _\delta}{1~\text{mas yr}^{-1}} \sin \Omega\right)\nonumber\\
    &\qquad- \left(\sqrt{1-e^2} \sin \omega+\cos \omega\right) \left(\frac{\mu_{\alpha*}}{1~\text{mas yr}^{-1}} \sin \Omega+\frac{\mu _\delta}{1~\text{mas yr}^{-1}} \cos \Omega\right)\Bigr]~\text{km s}^{-1}
\end{align}

\begin{align}
    \frac{d\Delta_\text{OP}}{dt}\approx&  d^{-1} a^{1/2} M^{1/2} \frac{1}{1-e} \Bigl\{\frac{1}{4} \sqrt{1-e^2} \sin 2 \omega \left[(e-2) \cos 2i +3 e-2\right]\nonumber\\
    &\qquad+(e-1) \cos^2\omega \left[(e+1) \cos^2 i +1\right]+\frac{1}{2} (e-1) \sin^2 \omega (-2 e+\cos 2i-1)\Bigr\}\\
    \approx& 1.14\times10^{-3} \left(\frac{d}{8~\text{kpc}}\right)^{-1} \left(\frac{a}{1000~\AU}\right)^{1/2} \left(\frac{M}{4\times10^6~\Msun}\right)^{1/2} \frac{1}{1-e} \Bigl\{\frac{1}{4} \sqrt{1-e^2} \sin 2 \omega \left[(e-2) \cos 2i +3 e-2\right]\nonumber\\
    &\qquad+(e-1) \cos^2\omega \left[(e+1) \cos^2 i +1\right]+\frac{1}{2} (e-1) \sin^2 \omega (-2 e+\cos 2i-1)\Bigr\}~\text{km s}^{-1}
\end{align}

\begin{align}
    \frac{d\Delta_\text{AOP}}{dt} \approx& r_\oplus d^{-1} a^{-1/2} M^{1/2} \frac{1}{1-e} \Bigl[\cos \omega \cos \Omega \left(1+\sqrt{1-e^2} \cos i\right)\nonumber\\
    &\qquad+\sin \omega \cos \Omega \left(\sqrt{1-e^2}+\cos i\right)+\cos \omega \sin \Omega \left(1-\sqrt{1-e^2} \cos i\right)+\sin \omega \sin \Omega \left(\sqrt{1-e^2}-\cos i\right)\Bigr]\nonumber\\
    +& a v_\oplus d^{-1} \Bigl[\left((e-1)-\sqrt{1-e^2} \cos i\right)\cos \omega (\cos \Omega + \sin \Omega)\nonumber\\
    &\qquad+\left(\sqrt{1-e^2}+(e-1) \cos i\right)\sin \omega (\cos \Omega + \sin \Omega)\Bigr]\\
    \approx& 1.14\times 10^{-6} \left(\frac{r_\oplus}{1~\AU}\right) \left(\frac{d}{8~\text{kpc}}\right)^{-1} \left(\frac{a}{1000~\text{AU}}\right)^{-1/2} \left(\frac{M}{4\times10^6~\Msun}\right)^{1/2} \frac{1}{1-e} \Bigl[\cos \omega \cos \Omega \left(1+\sqrt{1-e^2} \cos i\right)\nonumber\\
    &\qquad+\sin \omega \cos \Omega \left(\sqrt{1-e^2}+\cos i\right)+\cos \omega \sin \Omega \left(1-\sqrt{1-e^2} \cos i\right)+\sin \omega \sin \Omega \left(\sqrt{1-e^2}-\cos i\right)\Bigr]\nonumber\\
    +& 1.82\times10^{-5}\left(\frac{a}{1000~\AU}\right) \left(\frac{v_\oplus}{30~\text{km s}^{-1}}\right)\left(\frac{d}{8~\text{kpc}}\right)^{-1} \Bigl[\left((e-1)-\sqrt{1-e^2} \cos i\right)\cos \omega (\cos \Omega + \sin \Omega)\nonumber\\
    &\qquad+\left(\sqrt{1-e^2}+(e-1) \cos i\right)\sin \omega (\cos \Omega + \sin \Omega)\Bigr]~\text{km s}^{-1}
\end{align}

\noindent (vi) Earth Roemer Effect
\begin{align}
    \frac{d\RO}{dt} \approx& -2\pi r_\oplus P_\oplus \Bigl\{1- \left[\left(\mu_{\alpha*} + \mu_\delta\right)^2 + \mu_\delta^2 \right]t^2 + \left(\mu_{\alpha*} + \mu_\delta\right) - \frac{1}{2}\left(\mu_{\alpha*} + \mu_\delta\right)\mu_\parallel t^2\Bigr\}\\
    \approx& -2\pi r_\oplus P_\oplus\\
    \approx& 29.81~\text{km s}^{-1}
\end{align}

\section{Glossary of Parameters}
\label{appendix:parameters}
This appendix serves as a glossary and quick-reference for the many symbols, equations, and parameters used in this paper. Table~\ref{table:params_symbols} lists the observed, free, and derived parameters for our model, as well as their accompanying equation numbers and their units. Table~\ref{table:delays-effects} lists the timing delays and spectroscopic effects discussed and derived in this paper, as well as their corresponding equation numbers.

\begin{deluxetable}{cCccch}[htb!]
\tablecaption{Table of Parameters and Symbols
\label{table:params_symbols}}
\tablehead{\colhead{Parameter} & \colhead{Symbol} & \colhead{Meaning} & \colhead{Equation} & \colhead{Units} & \nocolhead{Dimensionless}\\
\colhead{Type} & \colhead{} & \colhead{} & \colhead{} & \colhead{} & \nocolhead{Units}}
\startdata
\multirow{3}{*}{Observed} & \alpha(t) & Right ascension (R.A.) & - & hms/decimal degr. & rad \\
 & \delta(t) & Declination (Decl.) & - & dms/decimal degr. & rad \\
 & v_z(t) & Line-of-sight velocity & \ref{eq:los-velocity} & km s$^{-1}$ & $c$ \\\hline
\multirow{7}{*}{\begin{tabular}{c}Free --\\System\end{tabular}} & M & Total mass of system & - & \Msun & $M_0$ \\
 & \alpha_i & Initial R.A. of binary barycenter & \ref{eq:alpha0} & hms/decimal degr. & rad \\
 & \delta_i & Initial Decl. of binary barycenter & \ref{eq:delta0} & dms/decimal degr. & rad \\ 
 & \mu_\alpha & R.A. proper motion & - & mas yr$^{-1}$ & rad $\tau_0^{-1}$ \\
 & \mu_\delta & Decl. proper motion & - & mas yr$^{-1}$ & rad $\tau_0^{-1}$ \\
 & \mu_\parallel & Line-of-sight proper motion & - & km s$^{-1}$ & $c$ \\
 & d & Distance from \SSB\ to \BB & - & kpc & $R_0$\\\hline
\multirow{7}{*}{\begin{tabular}{c}Free --\\Stellar\end{tabular}} & P & Period & - & s & $\tau_0$ \\
 & q & Mass ratio, $m_2 < m_1$  & \ref{eq:q} & - & - \\
 & e_R & Radial eccentricity & - & - & - \\
 & i & Inclination & - & \degr & rad\\
 & \Omega & Argument of ascending node & - & \degr & rad\\
 & \omega_0 & Initial argument of periapsis & - & \degr & rad\\
 & t_0 & Epoch of position  & - & yr & $\tau_0$\\\hline
\multirow{19}{*}{Derived} & \mu & Reduced mass & \ref{eq:mu} & \Msun & $M_0$ \\
 & \nu & Dimensionless reduced mass & \ref{eq:nu} & - & - \\
 & u & Eccentric anomaly & \ref{eq:kepler} & rad & rad \\
 & n & Mean motion & \ref{eq:n} & s\textsuperscript{-1} & $\tau_0^{-1}$ \\
 & \omega(u) & Time-dependent argument of periapsis & \ref{eq:omega_tdep} & \degr & rad\\
 & E & Energy of system (per mass) & \ref{eq:E} & $c^2$ & $E_0$ \\
 & J & Angular momentum of system (per mass) & \ref{eq:J} & km$^2$ s$^{-1}$ & $J_0$ \\
 & K & GR correction to $J$ & \ref{eq:K} & - & - \\
 & a_R & Semimajor axis & \ref{eq:aR} & AU & $R_0$ \\
 & a_{r_1},~a_{r_2} & Semimajor axes ($m_1$ and $m_2$) & \ref{eq:ar1}, \ref{eq:ar2} & AU & $R_0$ \\
 & e_t & Time eccentricity & \ref{eq:et} & - & - \\
 & e_\theta & Angular eccentricity & \ref{eq:etheta} & - & - \\  
 & e_{r_1},~e_{r_2} & Radial eccentricities ($m_1$ and $m_2$) & \ref{eq:er1}, \ref{eq:er2} & - & - \\
 & r_1(t),~r_2(t) & Radii at time $t$ ($m_1$ and $m_2$) & \ref{eq:r1}, \ref{eq:r2} & AU & $R_0$ \\
 & \theta(t) & Position angle at time $t$ & \ref{eq:theta} & rad & rad \\
 & \Delta\theta & Position angle precession & \ref{eq:precess} & rad & rad\\
  & \alpha_0(t) & R.A. of binary barycenter & \ref{eq:alpha0} & hms/decimal degr. & rad \\
 & \delta_0(t) & Decl. of binary barycenter & \ref{eq:delta0} & dms/decimal degr. & rad \\ 
  & \mu_\perp & Transverse proper motion & \ref{eq:muPerp} & mas yr$^{-1}$ & rad $\tau_0^{-1}$
\enddata
\end{deluxetable}

\begin{deluxetable}{Ccc}[htb!]
\tablecaption{Table of Timing Delay and Effects
\label{table:delays-effects}}
\tablehead{\colhead{Symbol} & \colhead{Meaning} & \colhead{Equation}}
\startdata
\RB & Binary Roemer Delay & \ref{eq:RB} \\
\EB & Binary Einstein Delay & \ref{eq:EB} \\
\SB & Binary Shapiro Delay & \ref{eq:SB} \\
\KB & Kopeikin Delay & \ref{eq:KB} \\
\SR & Viewing Angle Geometry Delay (\KB) & \ref{eq:SR} \\
\AOP & Annual Orbital Parallax Delay (\KB) & \ref{eq:AOP} \\
\OP & Orbital Parallax Delay (\KB) & \ref{eq:OP} \\
\RO & Earth Roemer Delay & \ref{eq:RO} \\\\
d\RB/dt & Binary Roemer Effect & \ref{eq:dRBdt} \\
v_\omega & Periastron Precession & \ref{eq:vomega} \\
d\EB/dt & Binary Einstein Effect & \ref{eq:dEBdt} \\
d\SB/dt & Binary Shapiro Effect & \ref{eq:dSBdt} \\
d\KB/dt & Kopeikin Effect & \ref{eq:dKBdt} \\
d\SR/dt & Viewing Angle Geometry (\KB) & \ref{eq:dSRdt} \\
d\AOP/dt & Annual Orbital Parallax (\KB) & \ref{eq:dAOPdt} \\
d\OP/dt & Orbital Parallax (\KB) & \ref{eq:dOPdt} \\
d\RO/dt & Earth Roemer Effect & \ref{eq:dROdt} \\
\enddata
\end{deluxetable}

\section{Parameter Values for S0-2 and S0-X}
\label{appendix:s02-s0x-params}
In Table~\ref{table:s02-s0x-vals}, we list the parameter values used for the S0-2 and S0-X stars in Section~\ref{sec:spec-effecs}. We have also set $M=3.985\times10^6$~\Msun\ and $d=7.971$~kpc \citep{2019Sci...365..664D}.

\begin{deluxetable}{CCC}[htb!]
\tablecaption{Parameter Values for S0-2 and S0-X Models 
\label{table:s02-s0x-vals}}
\tablehead{\colhead{Parameter} & \colhead{S0-2\tablenotemark{a}} & \colhead{S0-X\tablenotemark{b}}}
\startdata
P & 16.041~\text{yr} & 9.9~\text{yr}\\
q & 3.286 \times 10^{-6} & 3.286 \times 10^{-6}\\
e_R & 0.886 & 0.976\\
i & 133.88\degr & 72.76\degr\\
\Omega & 227.40\degr & -57.39\degr\\
\omega_0 & 66.03\degr & 42.62\degr\\
t_0 & 2018.3765~\text{yr} & 2003.33~\text{yr}\\
\enddata
\tablenotetext{a}{S0-2 values from \citet{2019Sci...365..664D}}
\tablenotetext{b}{S0-X values based on \citet{2020ApJ...889...61P}}
\end{deluxetable}

\bibliography{references}{}

\begin{thebibliography}{}
\expandafter\ifx\csname natexlab\endcsname\relax\def\natexlab#1{#1}\fi
\providecommand{\url}[1]{\href{#1}{#1}}
\providecommand{\dodoi}[1]{doi:~\href{http://doi.org/#1}{\nolinkurl{#1}}}
\providecommand{\doeprint}[1]{\href{http://ascl.net/#1}{\nolinkurl{http://ascl.net/#1}}}
\providecommand{\doarXiv}[1]{\href{https://arxiv.org/abs/#1}{\nolinkurl{https://arxiv.org/abs/#1}}}

\bibitem[{{Ang{\'e}lil} \& {Saha}(2014)}]{2014MNRAS.444.3780A}
{Ang{\'e}lil}, R., \& {Saha}, P. 2014, \mnras, 444, 3780,
  \dodoi{10.1093/mnras/stu1686}

\bibitem[{{Ang{\'e}lil} {et~al.}(2010){Ang{\'e}lil}, {Saha}, \&
  {Merritt}}]{2010ApJ...720.1303A}
{Ang{\'e}lil}, R., {Saha}, P., \& {Merritt}, D. 2010, \apj, 720, 1303,
  \dodoi{10.1088/0004-637X/720/2/1303}

\bibitem[{{Backer}(1994)}]{1994ASIC..445..403B}
{Backer}, D.~C. 1994, in NATO Advanced Study Institute (ASI) Series C, Vol.
  445, The Nuclei of Normal Galaxies: Lessons from the Galactic Center, ed.
  R.~{Genzel} \& A.~I. {Harris}, 403

\bibitem[{{Blanchet} {et~al.}(1998){Blanchet}, {Faye}, \&
  {Ponsot}}]{1998PhRvD..58l4002B}
{Blanchet}, L., {Faye}, G., \& {Ponsot}, B. 1998, \prd, 58, 124002,
  \dodoi{10.1103/PhysRevD.58.124002}

\bibitem[{{Boehle} {et~al.}(2016){Boehle}, {Ghez}, {Sch{\"o}del}, {Meyer},
  {Yelda}, {Albers}, {Martinez}, {Becklin}, {Do}, {Lu}, {Matthews}, {Morris},
  {Sitarski}, \& {Witzel}}]{2016ApJ...830...17B}
{Boehle}, A., {Ghez}, A.~M., {Sch{\"o}del}, R., {et~al.} 2016, \apj, 830, 17,
  \dodoi{10.3847/0004-637X/830/1/17}

\bibitem[{{Bozza} \& {Mancini}(2012)}]{2012ApJ...753...56B}
{Bozza}, V., \& {Mancini}, L. 2012, \apj, 753, 56,
  \dodoi{10.1088/0004-637X/753/1/56}

\bibitem[{{Damour} \& {Deruelle}(1985)}]{DD1}
{Damour}, T., \& {Deruelle}, N. 1985, Ann. Inst. Henri Poincar{\'e} Phys.
  Th{\'e}or, 43, 107

\bibitem[{{Damour} \& {Deruelle}(1986)}]{DD2}
---. 1986, Ann. Inst. Henri Poincar{\'e} Phys. Th{\'e}or, 44, 263

\bibitem[{{Damour} \& {Schafer}(1988)}]{1988NCimB.101..127D}
{Damour}, T., \& {Schafer}, G. 1988, Nuovo Cimento B Serie, 101B, 127,
  \dodoi{10.1007/BF02828697}

\bibitem[{{Do} {et~al.}(2013){Do}, {Martinez}, {Yelda}, {Ghez}, {Bullock},
  {Kaplinghat}, {Lu}, {Peter}, \& {Phifer}}]{2013ApJ...779L...6D}
{Do}, T., {Martinez}, G.~D., {Yelda}, S., {et~al.} 2013, \apjl, 779, L6,
  \dodoi{10.1088/2041-8205/779/1/L6}

\bibitem[{{Do} {et~al.}(2019){Do}, {Hees}, {Ghez}, {Martinez}, {Chu}, {Jia},
  {Sakai}, {Lu}, {Gautam}, {O'Neil}, {Becklin}, {Morris}, {Matthews},
  {Nishiyama}, {Campbell}, {Chappell}, {Chen}, {Ciurlo}, {Dehghanfar},
  {Gallego-Cano}, {Kerzendorf}, {Lyke}, {Naoz}, {Saida}, {Sch{\"o}del},
  {Takahashi}, {Takamori}, {Witzel}, \& {Wizinowich}}]{2019Sci...365..664D}
{Do}, T., {Hees}, A., {Ghez}, A., {et~al.} 2019, Science, 365, 664,
  \dodoi{10.1126/science.aav8137}

\bibitem[{{Downes} \& {Martin}(1971)}]{1971Natur.233..112D}
{Downes}, D., \& {Martin}, A.~H.~M. 1971, \nat, 233, 112,
  \dodoi{10.1038/233112a0}

\bibitem[{{Eckart} \& {Genzel}(1996)}]{1996Natur.383..415E}
{Eckart}, A., \& {Genzel}, R. 1996, \nat, 383, 415, \dodoi{10.1038/383415a0}

\bibitem[{{Eckart} \& {Genzel}(1997)}]{1997MNRAS.284..576E}
---. 1997, \mnras, 284, 576, \dodoi{10.1093/mnras/284.3.576}

\bibitem[{{Eckart} \& {Genzel}(1999)}]{1999JApA...20..187E}
---. 1999, Journal of Astrophysics and Astronomy, 20, 187,
  \dodoi{10.1007/BF02702351}

\bibitem[{{Eckart} {et~al.}(1999){Eckart}, {Ott}, \&
  {Genzel}}]{1999A&A...352L..22E}
{Eckart}, A., {Ott}, T., \& {Genzel}, R. 1999, \aap, 352, L22.
\newblock \doarXiv{astro-ph/9911011}

\bibitem[{{Edwards} {et~al.}(2006){Edwards}, {Hobbs}, \&
  {Manchester}}]{Tempo2006}
{Edwards}, R.~T., {Hobbs}, G.~B., \& {Manchester}, R.~N. 2006, \mnras, 372,
  1549, \dodoi{10.1111/j.1365-2966.2006.10870.x}

\bibitem[{{Fang} \& {Chen}(2021)}]{2021PhRvD.103f3041F}
{Fang}, Y., \& {Chen}, X. 2021, \prd, 103, 063041,
  \dodoi{10.1103/PhysRevD.103.063041}

\bibitem[{{Genzel} \& {Eckart}(1998)}]{1998IAUS..184..421G}
{Genzel}, R., \& {Eckart}, A. 1998, in The Central Regions of the Galaxy and
  Galaxies, ed. Y.~{Sofue}, Vol. 184, 421

\bibitem[{{Genzel} {et~al.}(1997){Genzel}, {Eckart}, {Ott}, \&
  {Eisenhauer}}]{1997MNRAS.291..219G}
{Genzel}, R., {Eckart}, A., {Ott}, T., \& {Eisenhauer}, F. 1997, \mnras, 291,
  219, \dodoi{10.1093/mnras/291.1.219}

\bibitem[{{Genzel} {et~al.}(1994){Genzel}, {Hollenbach}, \&
  {Townes}}]{1994RPPh...57..417G}
{Genzel}, R., {Hollenbach}, D., \& {Townes}, C.~H. 1994, Reports on Progress in
  Physics, 57, 417, \dodoi{10.1088/0034-4885/57/5/001}

\bibitem[{{Genzel} {et~al.}(1996){Genzel}, {Thatte}, {Krabbe}, {Kroker}, \&
  {Tacconi-Garman}}]{1996ApJ...472..153G}
{Genzel}, R., {Thatte}, N., {Krabbe}, A., {Kroker}, H., \& {Tacconi-Garman},
  L.~E. 1996, \apj, 472, 153, \dodoi{10.1086/178051}

\bibitem[{{Ghez} {et~al.}(1998){Ghez}, {Klein}, {Morris}, \&
  {Becklin}}]{1998ApJ...509..678G}
{Ghez}, A.~M., {Klein}, B.~L., {Morris}, M., \& {Becklin}, E.~E. 1998, \apj,
  509, 678, \dodoi{10.1086/306528}

\bibitem[{{Ghez} {et~al.}(2000){Ghez}, {Morris}, {Becklin}, {Tanner}, \&
  {Kremenek}}]{2000Natur.407..349G}
{Ghez}, A.~M., {Morris}, M., {Becklin}, E.~E., {Tanner}, A., \& {Kremenek}, T.
  2000, \nat, 407, 349, \dodoi{10.1038/35030032}

\bibitem[{{Ghez} {et~al.}(2003){Ghez}, {Duch{\^e}ne}, {Matthews}, {Hornstein},
  {Tanner}, {Larkin}, {Morris}, {Becklin}, {Salim}, {Kremenek}, {Thompson},
  {Soifer}, {Neugebauer}, \& {McLean}}]{2003ApJ...586L.127G}
{Ghez}, A.~M., {Duch{\^e}ne}, G., {Matthews}, K., {et~al.} 2003, \apjl, 586,
  L127, \dodoi{10.1086/374804}

\bibitem[{{Ghez} {et~al.}(2008){Ghez}, {Salim}, {Weinberg}, {Lu}, {Do}, {Dunn},
  {Matthews}, {Morris}, {Yelda}, {Becklin}, {Kremenek}, {Milosavljevic}, \&
  {Naiman}}]{2008ApJ...689.1044G}
{Ghez}, A.~M., {Salim}, S., {Weinberg}, N.~N., {et~al.} 2008, \apj, 689, 1044,
  \dodoi{10.1086/592738}

\bibitem[{{Gillessen} {et~al.}(2009){Gillessen}, {Eisenhauer}, {Fritz},
  {Bartko}, {Dodds-Eden}, {Pfuhl}, {Ott}, \& {Genzel}}]{2009ApJ...707L.114G}
{Gillessen}, S., {Eisenhauer}, F., {Fritz}, T.~K., {et~al.} 2009, \apjl, 707,
  L114, \dodoi{10.1088/0004-637X/707/2/L114}

\bibitem[{{Gillessen} {et~al.}(2017){Gillessen}, {Plewa}, {Eisenhauer}, {Sari},
  {Waisberg}, {Habibi}, {Pfuhl}, {George}, {Dexter}, {von Fellenberg}, {Ott},
  \& {Genzel}}]{2017ApJ...837...30G}
{Gillessen}, S., {Plewa}, P.~M., {Eisenhauer}, F., {et~al.} 2017, \apj, 837,
  30, \dodoi{10.3847/1538-4357/aa5c41}

\bibitem[{{Gravity Collaboration} {et~al.}(2018){Gravity Collaboration},
  {Abuter}, {Amorim}, {Anugu}, {Baub{\"o}ck}, {Benisty}, {Berger}, {Blind},
  {Bonnet}, {Brandner}, {Buron}, {Collin}, {Chapron}, {Cl{\'e}net}, {Coud{\'e}
  Du Foresto}, {de Zeeuw}, {Deen}, {Delplancke-Str{\"o}bele}, {Dembet},
  {Dexter}, {Duvert}, {Eckart}, {Eisenhauer}, {Finger}, {F{\"o}rster
  Schreiber}, {F{\'e}dou}, {Garcia}, {Garcia Lopez}, {Gao}, {Gendron},
  {Genzel}, {Gillessen}, {Gordo}, {Habibi}, {Haubois}, {Haug}, {Hau{\ss}mann},
  {Henning}, {Hippler}, {Horrobin}, {Hubert}, {Hubin}, {Jimenez Rosales},
  {Jochum}, {Jocou}, {Kaufer}, {Kellner}, {Kendrew}, {Kervella}, {Kok},
  {Kulas}, {Lacour}, {Lapeyr{\`e}re}, {Lazareff}, {Le Bouquin}, {L{\'e}na},
  {Lippa}, {Lenzen}, {M{\'e}rand}, {M{\"u}ler}, {Neumann}, {Ott}, {Palanca},
  {Paumard}, {Pasquini}, {Perraut}, {Perrin}, {Pfuhl}, {Plewa}, {Rabien},
  {Ram{\'\i}rez}, {Ramos}, {Rau}, {Rodr{\'\i}guez-Coira}, {Rohloff}, {Rousset},
  {Sanchez-Bermudez}, {Scheithauer}, {Sch{\"o}ller}, {Schuler}, {Spyromilio},
  {Straub}, {Straubmeier}, {Sturm}, {Tacconi}, {Tristram}, {Vincent}, {von
  Fellenberg}, {Wank}, {Waisberg}, {Widmann}, {Wieprecht}, {Wiest},
  {Wiezorrek}, {Woillez}, {Yazici}, {Ziegler}, \& {Zins}}]{2018A&A...615L..15G}
{Gravity Collaboration}, {Abuter}, R., {Amorim}, A., {et~al.} 2018, \aap, 615,
  L15, \dodoi{10.1051/0004-6361/201833718}

\bibitem[{{Gravity Collaboration} {et~al.}(2019){Gravity Collaboration},
  {Abuter}, {Amorim}, {Baub{\"o}ck}, {Berger}, {Bonnet}, {Brandner},
  {Cl{\'e}net}, {Coud{\'e} Du Foresto}, {de Zeeuw}, {Dexter}, {Duvert},
  {Eckart}, {Eisenhauer}, {F{\"o}rster Schreiber}, {Garcia}, {Gao}, {Gendron},
  {Genzel}, {Gerhard}, {Gillessen}, {Habibi}, {Haubois}, {Henning}, {Hippler},
  {Horrobin}, {Jim{\'e}nez-Rosales}, {Jocou}, {Kervella}, {Lacour},
  {Lapeyr{\`e}re}, {Le Bouquin}, {L{\'e}na}, {Ott}, {Paumard}, {Perraut},
  {Perrin}, {Pfuhl}, {Rabien}, {Rodriguez Coira}, {Rousset}, {Scheithauer},
  {Sternberg}, {Straub}, {Straubmeier}, {Sturm}, {Tacconi}, {Vincent}, {von
  Fellenberg}, {Waisberg}, {Widmann}, {Wieprecht}, {Wiezorrek}, {Woillez}, \&
  {Yazici}}]{2019A&A...625L..10G}
---. 2019, \aap, 625, L10, \dodoi{10.1051/0004-6361/201935656}

\bibitem[{{Gravity Collaboration} {et~al.}(2020){Gravity Collaboration},
  {Abuter}, {Amorim}, {Baub{\"o}ck}, {Berger}, {Bonnet}, {Brandner}, {Cardoso},
  {Cl{\'e}net}, {de Zeeuw}, {Dexter}, {Eckart}, {Eisenhauer}, {F{\"o}rster
  Schreiber}, {Garcia}, {Gao}, {Gendron}, {Genzel}, {Gillessen}, {Habibi},
  {Haubois}, {Henning}, {Hippler}, {Horrobin}, {Jim{\'e}nez-Rosales}, {Jochum},
  {Jocou}, {Kaufer}, {Kervella}, {Lacour}, {Lapeyr{\`e}re}, {Le Bouquin},
  {L{\'e}na}, {Nowak}, {Ott}, {Paumard}, {Perraut}, {Perrin}, {Pfuhl},
  {Rodr{\'\i}guez-Coira}, {Shangguan}, {Scheithauer}, {Stadler}, {Straub},
  {Straubmeier}, {Sturm}, {Tacconi}, {Vincent}, {von Fellenberg}, {Waisberg},
  {Widmann}, {Wieprecht}, {Wiezorrek}, {Woillez}, {Yazici}, \&
  {Zins}}]{2020A&A...636L...5G}
---. 2020, \aap, 636, L5, \dodoi{10.1051/0004-6361/202037813}

\bibitem[{{Gravity Collaboration} {et~al.}(2021){Gravity Collaboration},
  {Abuter}, {Amorim}, {Baub{\"o}ck}, {Berger}, {Bonnet}, {Brandner},
  {Cl{\'e}net}, {Davies}, {de Zeeuw}, {Dexter}, {Dallilar}, {Drescher},
  {Eckart}, {Eisenhauer}, {F{\"o}rster Schreiber}, {Garcia}, {Gao}, {Gendron},
  {Genzel}, {Gillessen}, {Habibi}, {Haubois}, {Hei{\ss}el}, {Henning},
  {Hippler}, {Horrobin}, {Jim{\'e}nez-Rosales}, {Jochum}, {Jocou}, {Kaufer},
  {Kervella}, {Lacour}, {Lapeyr{\`e}re}, {Le Bouquin}, {L{\'e}na}, {Lutz},
  {Nowak}, {Ott}, {Paumard}, {Perraut}, {Perrin}, {Pfuhl}, {Rabien},
  {Rodr{\'\i}guez-Coira}, {Shangguan}, {Shimizu}, {Scheithauer}, {Stadler},
  {Straub}, {Straubmeier}, {Sturm}, {Tacconi}, {Vincent}, {von Fellenberg},
  {Waisberg}, {Widmann}, {Wieprecht}, {Wiezorrek}, {Woillez}, {Yazici},
  {Young}, \& {Zins}}]{2021A&A...647A..59G}
---. 2021, \aap, 647, A59, \dodoi{10.1051/0004-6361/202040208}

\bibitem[{{Gravity Collaboration} {et~al.}(2022{\natexlab{a}}){Gravity
  Collaboration}, {Abuter}, {Aimar}, {Amorim}, {Ball}, {Baub{\"o}ck}, {Berger},
  {Bonnet}, {Bourdarot}, {Brandner}, {Cardoso}, {Cl{\'e}net}, {Dallilar},
  {Davies}, {de Zeeuw}, {Dexter}, {Drescher}, {Eisenhauer}, {F{\"o}rster
  Schreiber}, {Foschi}, {Garcia}, {Gao}, {Gendron}, {Genzel}, {Gillessen},
  {Habibi}, {Haubois}, {Hei{\ss}el}, {Henning}, {Hippler}, {Horrobin},
  {Jochum}, {Jocou}, {Kaufer}, {Kervella}, {Lacour}, {Lapeyr{\`e}re}, {Le
  Bouquin}, {L{\'e}na}, {Lutz}, {Ott}, {Paumard}, {Perraut}, {Perrin}, {Pfuhl},
  {Rabien}, {Shangguan}, {Shimizu}, {Scheithauer}, {Stadler}, {Stephens},
  {Straub}, {Straubmeier}, {Sturm}, {Tacconi}, {Tristram}, {Vincent}, {von
  Fellenberg}, {Widmann}, {Wieprecht}, {Wiezorrek}, {Woillez}, {Yazici}, \&
  {Young}}]{2022A&A...657L..12G}
{Gravity Collaboration}, {Abuter}, R., {Aimar}, N., {et~al.}
  2022{\natexlab{a}}, \aap, 657, L12, \dodoi{10.1051/0004-6361/202142465}

\bibitem[{{Gravity Collaboration} {et~al.}(2022{\natexlab{b}}){Gravity
  Collaboration}, {Abuter}, {Aimar}, {Amorim}, {Arras}, {Baub{\"o}ck},
  {Berger}, {Bonnet}, {Brandner}, {Bourdarot}, {Cardoso}, {Cl{\'e}net},
  {Davies}, {de Zeeuw}, {Dexter}, {Dallilar}, {Drescher}, {Eisenhauer},
  {En{\ss}lin}, {F{\"o}rster Schreiber}, {Garcia}, {Gao}, {Gendron}, {Genzel},
  {Gillessen}, {Habibi}, {Haubois}, {Hei{\ss}el}, {Henning}, {Hippler},
  {Horrobin}, {Jim{\'e}nez-Rosales}, {Jochum}, {Jocou}, {Kaufer}, {Kervella},
  {Lacour}, {Lapeyr{\`e}re}, {Le Bouquin}, {L{\'e}na}, {Lutz}, {Mang}, {Nowak},
  {Ott}, {Paumard}, {Perraut}, {Perrin}, {Pfuhl}, {Rabien}, {Shangguan},
  {Shimizu}, {Scheithauer}, {Stadler}, {Straub}, {Straubmeier}, {Sturm},
  {Tacconi}, {Tristram}, {Vincent}, {von Fellenberg}, {Waisberg}, {Widmann},
  {Wieprecht}, {Wiezorrek}, {Woillez}, {Yazici}, {Young}, \&
  {Zins}}]{2022A&A...657A..82G}
---. 2022{\natexlab{b}}, \aap, 657, A82, \dodoi{10.1051/0004-6361/202142459}

\bibitem[{{Grould} {et~al.}(2017){Grould}, {Vincent}, {Paumard}, \&
  {Perrin}}]{2017A&A...608A..60G}
{Grould}, M., {Vincent}, F.~H., {Paumard}, T., \& {Perrin}, G. 2017, \aap, 608,
  A60, \dodoi{10.1051/0004-6361/201731148}

\bibitem[{{Hei{\ss}el} {et~al.}(2022){Hei{\ss}el}, {Paumard}, {Perrin}, \&
  {Vincent}}]{2022A&A...660A..13H}
{Hei{\ss}el}, G., {Paumard}, T., {Perrin}, G., \& {Vincent}, F. 2022, \aap,
  660, A13, \dodoi{10.1051/0004-6361/202142114}

\bibitem[{{Jiang} \& {Lin}(1985)}]{1985AmJPh..53..694J}
{Jiang}, H.~X., \& {Lin}, J.~Y. 1985, American Journal of Physics, 53, 694,
  \dodoi{10.1119/1.14287}

\bibitem[{{Kormendy} \& {Ho}(2013)}]{2013ARA&A..51..511K}
{Kormendy}, J., \& {Ho}, L.~C. 2013, \araa, 51, 511,
  \dodoi{10.1146/annurev-astro-082708-101811}

\bibitem[{{Kormendy} \& {Richstone}(1995)}]{1995ARA&A..33..581K}
{Kormendy}, J., \& {Richstone}, D. 1995, \araa, 33, 581,
  \dodoi{10.1146/annurev.aa.33.090195.003053}

\bibitem[{{Lo}(1989)}]{1989IAUS..136..527L}
{Lo}, K.~Y. 1989, in The Center of the Galaxy, ed. M.~{Morris}, Vol. 136, 527

\bibitem[{{Lo} {et~al.}(1993){Lo}, {Backer}, {Kellermann}, {Reid}, {Zhao},
  {Goss}, \& {Moran}}]{1993Natur.362...38L}
{Lo}, K.~Y., {Backer}, D.~C., {Kellermann}, K.~I., {et~al.} 1993, \nat, 362,
  38, \dodoi{10.1038/362038a0}

\bibitem[{{Luzum} \& {Petit}(2015)}]{2015HiA....16..227L}
{Luzum}, B., \& {Petit}, G. 2015, Highlights of Astronomy, 16, 227,
  \dodoi{10.1017/S1743921314005535}

\bibitem[{{Marconi} {et~al.}(2021){Marconi}, {Abreu}, {Adibekyan}, {Aliverti},
  {Allende Prieto}, {Amado}, {Amate}, {Artigau}, {Augusto}, {Barros},
  {Becerril}, {Benneke}, {Bergin}, {Berio}, {Bezawada}, {Boisse}, {Bonfils},
  {Bouchy}, {Broeg}, {Cabral}, {Calvo-Ortega}, {Canto Martins}, {Chazelas},
  {Chiavassa}, {Christensen}, {Cirami}, {Coretti}, {Covino}, {Cresci},
  {Cristiani}, {Cunha Parro}, {Cupani}, {de Castro Le{\~a}o}, {Renan de
  Medeiros}, {Furlande Souza}, {Di Marcantonio}, {Di Varano}, {D'Odorico},
  {Doyon}, {Drass}, {Figueira}, {Belen Fragoso}, {Uldall Fynbo}, {Gallo},
  {Genoni}, {Gonz{\'a}lez Hern{\'a}ndez}, {Haehnelt}, {Hlavacek-Larrondo},
  {Hughes}, {Huke}, {Humphrey}, {Kjeldsen}, {Korn}, {Kouach}, {Landoni},
  {Liske}, {Lovis}, {Lunney}, {Maiolino}, {Malo}, {Marquart}, {Martins},
  {Mason}, {Molaro}, {Monnier}, {Monteiro}, {Mordasini}, {Morris},
  {Mucciarelli}, {Murray}, {Niedzielski}, {Nunes}, {Oliva}, {Origlia},
  {Pall{\'e}}, {Pariani}, {Parr-Burman}, {Pe{\~n}ate}, {Pepe}, {Pinna},
  {Piskunov}, {Rasilla Pi{\~n}eiro}, {Rebolo}, {Rees}, {Reiners}, {Riva},
  {Romano}, {Rousseau}, {Sanna}, {Santos}, {Sarajlic}, {Shen}, {Sortino},
  {Sosnowska}, {Sousa}, {Stempels}, {Strassmeier}, {Tenegi}, {Tozzi}, {Udry},
  {Valenziano}, {Vanzi}, {Weber}, {Woche}, {Xompero}, {Zackrisson}, \&
  {Zapatero Osorio}}]{2021Msngr.182...27M}
{Marconi}, A., {Abreu}, M., {Adibekyan}, V., {et~al.} 2021, The Messenger, 182,
  27, \dodoi{10.18727/0722-6691/5219}

\bibitem[{{Martin} {et~al.}(2018){Martin}, {Fitzgerald}, {McLean}, {Doppmann},
  {Kassis}, {Aliado}, {Canfield}, {Johnson}, {Kress}, {Lanclos}, {Magnone},
  {Sohn}, {Wang}, \& {Weiss}}]{2018SPIE10702E..0AM}
{Martin}, E.~C., {Fitzgerald}, M.~P., {McLean}, I.~S., {et~al.} 2018, in
  Society of Photo-Optical Instrumentation Engineers (SPIE) Conference Series,
  Vol. 10702, Ground-based and Airborne Instrumentation for Astronomy VII, ed.
  C.~J. {Evans}, L.~{Simard}, \& H.~{Takami}, 107020A,
  \dodoi{10.1117/12.2312266}

\bibitem[{{Martins} {et~al.}(2008){Martins}, {Gillessen}, {Eisenhauer},
  {Genzel}, {Ott}, \& {Trippe}}]{2008ApJ...672L.119M}
{Martins}, F., {Gillessen}, S., {Eisenhauer}, F., {et~al.} 2008, \apjl, 672,
  L119, \dodoi{10.1086/526768}

\bibitem[{{Mawet} {et~al.}(2019){Mawet}, {Fitzgerald}, {Konopacky}, {Beichman},
  {Jovanovic}, {Dekany}, {Hover}, {Chisholm}, {Ciardi}, {Artigau}, {Banyal},
  {Beatty}, {Benneke}, {Blake}, {Burgasser}, {Canalizo}, {Chen}, {Do},
  {Doppmann}, {Doyon}, {Dressing}, {Fang}, {Greene}, {Hillenbrand}, {Howard},
  {Kane}, {Kataria}, {Kempton}, {Knutson}, {Kotani}, {Lafreni{\`e}re}, {Liu},
  {Nishiyama}, {Pandey}, {Plavchan}, {Prato}, {Rajaguru}, {Robertson}, {Salyk},
  {Sato}, {Schlawin}, {Sengupta}, {Sivarani}, {Skidmore}, {Tamura}, {Terada},
  {Vasisht}, {Wang}, \& {Zhang}}]{2019BAAS...51g.134M}
{Mawet}, D., {Fitzgerald}, M., {Konopacky}, Q., {et~al.} 2019, in Bulletin of
  the American Astronomical Society, Vol.~51, 134.
\newblock \doarXiv{1908.03623}

\bibitem[{{McGee} \& {Bolton}(1954)}]{1954Natur.173..985M}
{McGee}, R.~X., \& {Bolton}, J.~G. 1954, \nat, 173, 985,
  \dodoi{10.1038/173985b0}

\bibitem[{{Merritt} {et~al.}(2010){Merritt}, {Alexander}, {Mikkola}, \&
  {Will}}]{2010PhRvD..81f2002M}
{Merritt}, D., {Alexander}, T., {Mikkola}, S., \& {Will}, C.~M. 2010, \prd, 81,
  062002, \dodoi{10.1103/PhysRevD.81.062002}

\bibitem[{{Nusser} \& {Broadhurst}(2004)}]{2004MNRAS.355L...6N}
{Nusser}, A., \& {Broadhurst}, T. 2004, \mnras, 355, L6,
  \dodoi{10.1111/j.1365-2966.2004.08484.x}

\bibitem[{{O'Neil} {et~al.}(2019){O'Neil}, {Martinez}, {Hees}, {Ghez}, {Do},
  {Witzel}, {Konopacky}, {Becklin}, {Chu}, {Lu}, {Matthews}, \&
  {Sakai}}]{2019AJ....158....4O}
{O'Neil}, K.~K., {Martinez}, G.~D., {Hees}, A., {et~al.} 2019, \aj, 158, 4,
  \dodoi{10.3847/1538-3881/ab1d66}

\bibitem[{{Pei{\ss}ker} {et~al.}(2020{\natexlab{a}}){Pei{\ss}ker}, {Eckart}, \&
  {Parsa}}]{2020ApJ...889...61P}
{Pei{\ss}ker}, F., {Eckart}, A., \& {Parsa}, M. 2020{\natexlab{a}}, \apj, 889,
  61, \dodoi{10.3847/1538-4357/ab5afd}

\bibitem[{{Pei{\ss}ker} {et~al.}(2020{\natexlab{b}}){Pei{\ss}ker}, {Eckart},
  {Zaja{\v{c}}ek}, {Ali}, \& {Parsa}}]{2020ApJ...899...50P}
{Pei{\ss}ker}, F., {Eckart}, A., {Zaja{\v{c}}ek}, M., {Ali}, B., \& {Parsa}, M.
  2020{\natexlab{b}}, \apj, 899, 50, \dodoi{10.3847/1538-4357/ab9c1c}

\bibitem[{{Psaltis} {et~al.}(2016){Psaltis}, {Wex}, \&
  {Kramer}}]{2016ApJ...818..121P}
{Psaltis}, D., {Wex}, N., \& {Kramer}, M. 2016, \apj, 818, 121,
  \dodoi{10.3847/0004-637X/818/2/121}

\bibitem[{{Rubilar} \& {Eckart}(2001)}]{2001A&A...374...95R}
{Rubilar}, G.~F., \& {Eckart}, A. 2001, \aap, 374, 95,
  \dodoi{10.1051/0004-6361:20010640}

\bibitem[{{Sch{\"o}del} {et~al.}(2002){Sch{\"o}del}, {Ott}, {Genzel},
  {Hofmann}, {Lehnert}, {Eckart}, {Mouawad}, {Alexander}, {Reid}, {Lenzen},
  {Hartung}, {Lacombe}, {Rouan}, {Gendron}, {Rousset}, {Lagrange}, {Brandner},
  {Ageorges}, {Lidman}, {Moorwood}, {Spyromilio}, {Hubin}, \&
  {Menten}}]{2002Natur.419..694S}
{Sch{\"o}del}, R., {Ott}, T., {Genzel}, R., {et~al.} 2002, \nat, 419, 694,
  \dodoi{10.1038/nature01121}

\bibitem[{{Thatte} {et~al.}(1998){Thatte}, {Tecza}, {Eisenhauer}, {Mengel},
  {Krabbe}, {Pak}, {Genzel}, {Bonaccini}, {Emsellem}, {Rigaut}, {Delabre}, \&
  {Monnet}}]{1998SPIE.3353..704T}
{Thatte}, N.~A., {Tecza}, M., {Eisenhauer}, F., {et~al.} 1998, in Society of
  Photo-Optical Instrumentation Engineers (SPIE) Conference Series, Vol. 3353,
  Adaptive Optical System Technologies, ed. D.~{Bonaccini} \& R.~K. {Tyson},
  704--715, \dodoi{10.1117/12.321638}

\bibitem[{{Waisberg} {et~al.}(2018){Waisberg}, {Dexter}, {Gillessen}, {Pfuhl},
  {Eisenhauer}, {Plewa}, {Baub{\"o}ck}, {Jimenez-Rosales}, {Habibi}, {Ott},
  {von Fellenberg}, {Gao}, {Widmann}, \& {Genzel}}]{2018MNRAS.476.3600W}
{Waisberg}, I., {Dexter}, J., {Gillessen}, S., {et~al.} 2018, \mnras, 476,
  3600, \dodoi{10.1093/mnras/sty476}

\bibitem[{{Weinberg} \& {Milosavljevic}(2004)}]{2004AAS...204.1702W}
{Weinberg}, N.~N., \& {Milosavljevic}, M. 2004, in American Astronomical
  Society Meeting Abstracts, Vol. 204, American Astronomical Society Meeting
  Abstracts \#204, 17.02

\bibitem[{{Wex}(1995)}]{1995CQGra..12..983W}
{Wex}, N. 1995, Classical and Quantum Gravity, 12, 983,
  \dodoi{10.1088/0264-9381/12/4/009}

\bibitem[{{Wex} \& {Kopeikin}(1999)}]{1999ApJ...514..388W}
{Wex}, N., \& {Kopeikin}, S.~M. 1999, \apj, 514, 388, \dodoi{10.1086/306933}

\bibitem[{{Will}(2008)}]{2008ApJ...674L..25W}
{Will}, C.~M. 2008, \apjl, 674, L25, \dodoi{10.1086/528847}

\end{thebibliography}
\bibliographystyle{aasjournal}
\end{document}